\documentclass{emulateapj}
\usepackage{graphicx}
\usepackage{natbib}
\usepackage{subfigure}
\usepackage{url}
\begin{document}

\title{Direct Spectral Detection: An Efficient Method to Detect and Characterize Binary Systems}

\author{Kevin Gullikson \altaffilmark{1}}
\author{Adam Kraus \altaffilmark{1}}
\author{Sarah Dodson-Robinson \altaffilmark{2}}
\author{Daniel Jaffe \altaffilmark{1}}
\author{Jeong-Eun Lee \altaffilmark{3}}
\author{Gregory N. Mace \altaffilmark{1}}
\author{Phillip MacQueen \altaffilmark{1}}
\author{Sunkyung Park \altaffilmark{3}}
\author{Andrew Riddle \altaffilmark{1}}

\altaffiltext{1}{University of Texas, Astronomy Department. 2515 Speedway, Stop C1400. Austin, TX 78712 \email{kgulliks@astro.as.utexas.edu}}
\altaffiltext{2}{University of Delaware, Department of Physics and Astronomy, 217 Sharp Lab, Newark, DE 19716}
\altaffiltext{3}{School of Space Research, Kyung Hee University, Yongin-shi, Kyungki-do 449-701, Korea}

\begin{abstract}
Young, intermediate-mass stars are experiencing renewed interest as targets for direct-imaging planet searches. However, these types of stars are part of multiple systems more often than not. Close stellar companions affect the formation and orbital evolution of any planets, and the properties of the companions can help constrain the binary formation mechanism. Unfortunately, close companions are difficult and expensive to detect with imaging techniques. In this paper, we describe the direct spectral detection method wherein a high-resolution spectrum of the primary is cross-correlated against a template for a companion star. Variants of this method have previously been used to search for stellar, brown dwarf, and even planetary companions. We show that the direct spectral detection method can detect companions as late as M-type orbiting A0 or earlier primary stars in a single epoch on small-aperture telescopes. In addition to estimating the detection limits, we determine the sources of uncertainty in characterizing the companion temperature, and find that large systematic biases can exist. After calibrating the systematic biases with synthetic binary star observations, we apply the method to a sample of 34 known binary systems with an A- or B-type primary star. We detect nine total companions, including four of the five known companions with literature temperatures between $4000$ K $ < T < 6000$ K, the temperature range for which our method is optimized. We additionally characterize the companion for the first time in two previously single-lined binary systems and one binary identified with speckle interferometry. This method provides an inexpensive way to use small-aperture telescopes to detect binary companions with moderate mass-ratios, and is competitive with high-resolution imaging techniques inside $\sim 100-200$ mas.


\end{abstract}

\maketitle

\section{Introduction}

\label{sec:intro}
There has recently been a revival of interest in the multiplicity properties of intermediate-mass stars,  spurred largely by the detection of planets orbiting nearby \mbox{$\sim2 M_{\odot}$} stars on both wide \citep[e.g.][]{Lagrange2010, Marois2008} and close \citep{Johnson2011} orbits. In this context, binary companions are contaminants; companions complicate radial velocity planet searches because they necessitate simultaneous modeling of both stellar motions \citep[e.g.][]{Bergmann2015}. Likewise, companions complicate direct imaging planet searches by requiring either extremely high-contrast instrumentation \citep{Thalmann2014} or specialized coronagraphs \citep{Crepp2010}. 

However, known binary stars are typically avoided in planet search programs for a more fundamental reason: the binary companion depletes or destroys the planet-forming disk. By combining a binary census of the $\sim 2$ Myr Taurus-Auriga star-forming region with a disk census of the same, \cite{Kraus2012} showed that close ($\lesssim 40$ AU) binaries are about 2-3 times less likely to host a protoplanetary disk, and so hasten disk dispersal. Even if a disk survives, it tends to be depleted in mass by a factor of $\sim 25$ for binary separations $\lesssim 30$ AU \citep{Harris2012}.

Multiplicity is an inevitable outcome of star formation, especially for more massive stars where multiplicity is more common \citep{Zinnecker2007}, and so is important to study in its own right. Beyond the overall multiplicity rate, the mass ratio, period, and eccentricity distributions of a binary star population can place important constraints on the mode of binary star formation. Specifically, a mass-ratio distribution that changes with physical separation could indicate that secondary stars are either forming through disk fragmentation \citep[e.g.][]{Kratter2006, Stamatellos2011}  or are accreting a significant amount of their mass from the primary star disk. \cite{DeRosa2014} recently completed an imaging survey of nearby A-type stars, and found preliminary evidence that the mass-ratio distribution does in fact become flatter for closer companions. There is no evidence of such a change for solar-type or later stars \citep{Meyer2013}.

It is difficult to detect low-mass companions very near an intrinsically bright primary star and even more difficult to characterize the companion. There are three commonly used techniques for binary star searches: direct imaging with adaptive optics systems,  interferometry, and radial velocity monitoring. Imaging can easily detect low-mass companions at wide apparent separations, but loses sensitivity as the on-sky distance from the primary star decreases \citep[see][for typical sensitivity curves]{DeRosa2014}. Interferometry can usually achieve smaller working angles than imaging, but cannot achieve as high contrast \citep[see e.g.][]{Aldoretta2015}.

 Radial velocity monitoring can easily find companions on very short-period orbits, but its sensitivity to low-mass companions drops as the physical separation increases. This is especially true for A- and B-type stars, where radial velocity precision is typically limited to $\sim 1 \rm km \  s^{-1}$ by their rotationally broadened lines. Worse, radial velocity monitoring techniques cannot characterize the companion unless the inclination is known or if the companion spectral lines are also visible. All three techniques have separation-dependent sensitivity, which introduces observational bias in any search for a parameter that changes with physical separation.
 
 One technique that is separation independent is to search directly for the composite spectrum of two stars. \citet{Burgasser2007} used single-epoch low-resolution spectroscopy to identify and characterize a brown dwarf binary system by fitting both spectra simultaneously. This method only works if the stars have a similar brightness but very different spectra, such that the spectral features from both components are visible and distinguishable. If, as in the case of binary systems with very large flux ratios, the companion spectrum is buried within the noise of the primary star, a different method is needed.
 
The direct spectral detection method (hereafter referred to as the DSD method), and variations thereof, has been used to search for binary companions to early B-stars \citep{Gullikson2013}, main sequence FGK-stars \citep{Kolbl2015}, young K-M stars \citep{Prato2002}, and even `Hot Jupiter' type planets \citep{Snellen2010, Brogi2012, deKok2013} orbiting FGK-stars. The method relies on the cross-correlation function (CCF) of a high-spectral-resolution spectrum of the primary star with a model spectrum for the expected companion. The CCF uses every pixel in the spectrum, and more importantly every spectral line in the secondary spectrum. A simple experiment with synthetic spectra containing increasing numbers of spectral lines (N) in noisy data shows that the CCF peak significance increases as $\sim\sqrt{N}$. For high-resolution cross-dispersed \'echelle spectra, this amplification can reach several factors of 10, allowing the detection \emph{and characterization} of a secondary spectrum where the individual lines are completely buried in noise. Since the DSD method uses a seeing-limited spectrum of the primary star, its sensitivity is independent of separation inside $\sim 1 ''$ and can make use of small telescopes to detect high-contrast companions.

In this paper, we describe the DSD method in detail, and use it to detect the secondary star in nine known binary systems. We describe the method in Section \ref{sec:method}. We describe the observations and data reduction in Section \ref{sec:obs}, then use the observations to estimate the accuracy with which we can measure the companion temperature in Section \ref{subsec:systematics} and the sensitivity of the method in Section \ref{subsec:sensitivity}. In Section \ref{sec:results} we use the DSD method to search for known companions, and discuss the results in Section \ref{sec:conclusions}.

\section{Direct Spectral Detection Method}
\label{sec:method}

All implementations of the DSD method use high-spectral-resolution and high signal-to-noise spectra, and search for companions with extreme flux ratios by cross-correlating the observed spectra with models for the expected companion. The main differences between the various implementations are the primary star and telluric line removal processes. The `Hot Jupiter' searches \citep[e.g.][]{Snellen2010} use known orbital phase information and a high degree of phase coverage to simultaneously estimate an empirical stellar and telluric spectrum with minimal contamination from the planet, while \cite{Kolbl2015} subtract a best-fit model spectrum for the primary star. Since \cite{Kolbl2015} use optical data, they do not attempt any telluric correction. Note that the approach of \cite{Kolbl2015} is conceptually similar to the todcor code \citep{Mazeh1994}, which is widely used to search for double-lined spectroscopic binary systems.

Unlike most previous work, we focus on not only detecting but accurately characterizing the companion. We additionally optimize our technique for detecting cool companions to rapidly rotating early-type stars, for which it is very difficult to detect the reflex motion of the primary star. We fit and remove the telluric absorption using the TelFit code \citep{Gullikson2014}, and estimate an empirical primary star spectrum with a Gaussian smoothing filter applied to the telluric-corrected data. We chose to use a smoothing filter over subtracting model spectra for two reasons: first, the model spectra are a poor representation of the data, especially at the high signal-to-noise ratios that we use, and so leave very large-scale features in the residual spectrum. Second, the smoothing filter will also remove any large-scale instrumental systematics in the spectrum. We use a smoothing filter with a  window size ($w$) set by

\begin{equation}
w = \frac{v\sin{i} \cdot f}{c} \cdot \frac{\lambda_0}{\Delta \lambda}
\label{eqn:smoothing}
\end{equation}
where $v\sin{i}$ is the literature rotational velocity of the star, $\lambda_0$ is the central wavelength of the \'echelle order, $\Delta \lambda$ is the wavelength spacing per pixel of the order, $c$ is the speed of light, and $f = 0.25$ is an empirically determined parameter to give a visually adequate fit. Typical window sizes ranged from 50 - 100 pixels.

We use the following subset of the Phoenix library of model spectra prepared by \cite{Husser2013_b} throughout this work:

\begin{itemize}
\item $\rm T_{eff} = 3000-7000$ K\footnote{We extend the grid to higher temperatures if the measured temperature (see Section \ref{subsec:systematics}) is near 7000 K}, in steps of 100 K
\item {[}Fe/H{]} = -0.5, 0.0, +0.5
\item $v\sin{i} = 1, 5, 10, 20, 30 \ \rm km \ s^{-1}$
\end{itemize}
Here, the $v\sin{i}$ is the rotational velocity of the secondary star. We account for the small influence that the smoothing kernel has on the companion spectrum by convolving the model with the same smoothing kernel used for the data, and subtracting the convolved model from the original. Treating the model spectrum in this way is more commonly known as unsharp masking. Finally, we cross-correlate every \'echelle order that does not have strong telluric residuals against the corresponding model spectrum, and combine the CCFs for each order using a simple average. The method is summarized below:

\begin{enumerate}
 \item Smooth the observed spectrum with a Gaussian smoothing kernel with width given by Equation \ref{eqn:smoothing}, and subtract the smoothed spectrum from the original
 \item Rotationally broaden the Phoenix model spectrum to the requested companion $v\sin{i}$
 \item Smooth the broadened spectrum to the instrumental resolution by convolving it with a gaussian kernel of appropriate width.
 \item Unsharp mask the broadened model spectrum
 \item Resample the processed model spectrum to the same wavelength spacing per pixel as the observed spectrum.
 \item Cross-correlate each \'echelle order against the corresponding processed model spectrum, and combine using a simple average.
\end{enumerate}

\section{Observations and Data Reduction}
\label{sec:obs}
We use three separate samples in this work. The first set, given in Table \ref{tab:earlycal}, contains  A- and B-type stars with the following published properties:

\begin{itemize}
\item Spectral type B0V - A9V (only main sequence)
\item $V < 6$
\item $v\sin{i} > 80 \rm km \ s^{-1}$
\item No known companions within $3 ''$, and no sign of a companion in our data.
\end{itemize}
The lower limit on $v\sin{i}$ in our sample ensures that the empirical primary star template is accurate and only minimally affects any companions.

The second dataset (Table \ref{tab:latecal}) contains F-M type stars which have a high-quality temperature estimate in the literature. We use the first two samples in Sections \ref{subsec:systematics} and \ref{subsec:sensitivity} to assess the accuracy of the temperature estimation using the DSD method and the sensitivity to companions of various temperatures.

Finally, we use the third dataset (Table \ref{tab:known}) to search for the spectral lines of the companion in several known binary systems. The third dataset has the same properties as the first, except that they have one \emph{and only one} known companion within $1 ''$. We further require that the literature data either puts no constraints on the companion temperature (as in the case of single-lined spectroscopic binaries) or that the companion has $T_{\rm eff} < 7500$ K. 

We estimate the expected companion temperature depending on whether it is part of a spectroscopic (Table \ref{tab:specdata}) or visual (Table \ref{tab:imagedata}) binary system. In the case of double-lined spectroscopic binaries, we use the ratio of the semi-amplitudes given in the 9th catalog of spectroscopic binary orbits \citep[SB9,][]{SB9} to estimate the mass ratio of the system. We convert the primary star spectral type from the Simbad database \citep{Simbad} to mass by interpolating Table 5 of \citet{Pecaut2013}. The mass ratio and primary mass gives an estimate of the companion mass, which we convert to temperature by interpolating the same table. Most of the directly-imaged binary systems do not have orbital information, so we use the magnitude difference published in the Washington Double Star catalog \citep[WDS, ][]{WDS}. We use the Simbad spectral type of the primary star and Table 5 of \citet{Pecaut2013} to estimate the primary star temperature ($T_1$) and radius ($R_1$). We then find the companion temperature that minimizes the following function for the companion temperature ($T_2$), given the observed magnitude difference ($\Delta m_{obs}$)
\begin{equation}
Q = (m(T_2, R_2) - m(T_1, R_1) - \Delta m_{obs})^2
\label{eqn:secteff}
\end{equation}
where $m(T, R)$ is the Vega magnitude of a star with temperature T and radius R. We use the pysynphot package\footnote{pysynphot is a python code package to perform synthetic photometry, and is available at this url: \url{https://pypi.python.org/pypi/pysynphot}} and a Kurucz model grid \citep{Castelli2003} to calculate $m(T, R)$, and assume the companion is on the main sequence to estimate its radius ($R_2$). For both spectroscopic and visual binary systems, we assume spectral type uncertainties of $\pm 1$ subtype on the primary stars, and propagate the uncertainties into uncertainty in the companion temperature. We include the binary system in the sample if the expected companion temperature is $< 7500$ K. 

We use the same set of instruments and settings for all observations throughout the three datasets. We use the CHIRON spectrograph \citep{CHIRON} on the 1.5m telescope at Cerro Tololo Inter-American Observatory for most southern targets. This spectrograph is an $R\equiv \lambda / \Delta \lambda = 80000$ cross-dispersed \'echelle spectrograph with wavelength coverage from 450 - 850 nm, and is fed by a $2.7''$ optical fiber. The data are automatically reduced with a standard CHIRON data reduction pipeline, but the pipeline leaves residuals of strong lines in adjacent orders. We therefore bias-correct, flat-field and extract the spectra with the optimum extraction technique \citep{Horne1986} using standard IRAF\footnote{IRAF is distributed by the National Optical Astronomy Observatories, which are operated by the Association of Universities for Research in Astronomy, Inc., under cooperative agreement with the National Science Foundation.} tasks, and use the wavelength calibration from the pipeline reduced spectra.

For the northern targets, we use a combination of the High Resolution Spectrograph \citep[HRS,][]{HRS} on the Hobby Eberly Telescope, and the Tull coud\'e \citep[TS23,][]{TS23} and IGRINS \citep{IGRINS} spectrographs, both on the 2.7m Harlan J. Smith Telescope. All three northern instruments are at McDonald Observatory. For the HRS, we use the $R = 60000$ setting with a $2''$ fiber, and with wavelength coverage from 410-780 nm. We bias-correct, flat-field, and extract the spectra using an IRAF pipeline. The HRS spectra are wavelength-calibrated using a Th-Ar lamp observed immediately before or after the science observations.

For the TS23, we use a $1.2''$ slit in combination with the E2 \'echelle grating (53 grooves/mm, blaze angle $65^{\circ}$), yielding a resolving power of $R=60000$ and a wavelength coverage from 375-1020 nm. We reduce the data using an IRAF pipeline very similar to the one we use for the HRS, and wavelength calibrate using a Th-Ar lamp observed immediately before the science observations.

IGRINS only has one setting with $R = 40000$. It has complete wavelength coverage from 1475-2480 nm, except for where telluric absorption is almost $100\%$ from 1810 - 1930 nm. Each star is observed in an ABBA nodding mode, and reduced using the standard IGRINS pipeline \citep{IGRINS_plp_v2}. The standard pipeline uses atmospheric OH emission lines as well as a Th-Ar calibration frame to calibrate the wavelengths; we further refine the wavelength solution using telluric absorption lines and the TelFit code.

\begin{figure}
        \centering
        \subfigure[TS23]{
            \label{fig:error:a}
            \includegraphics[scale=0.45]{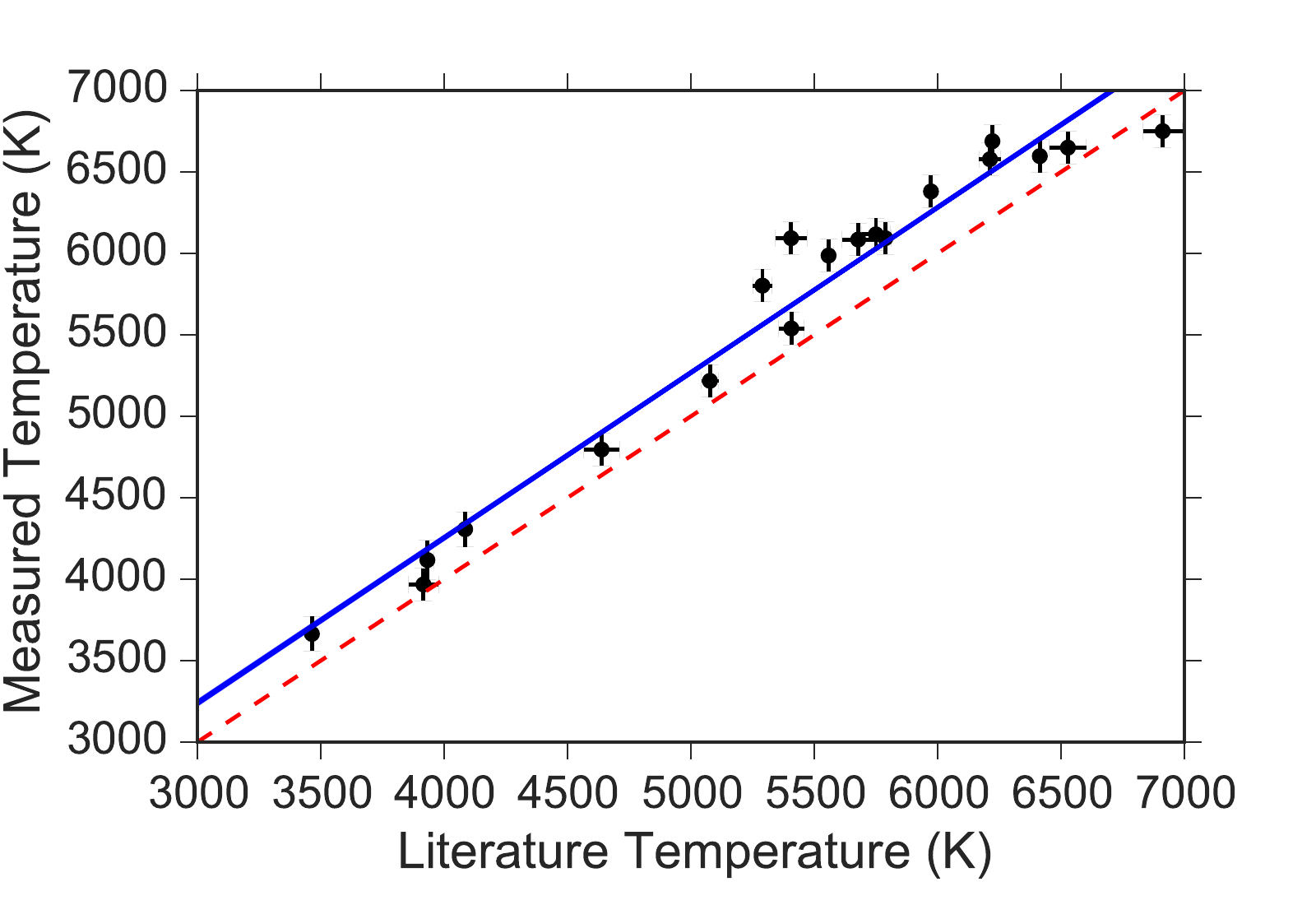}
         }
         \subfigure[HRS]{
             \label{fig:error:b}
             \includegraphics[scale=0.45]{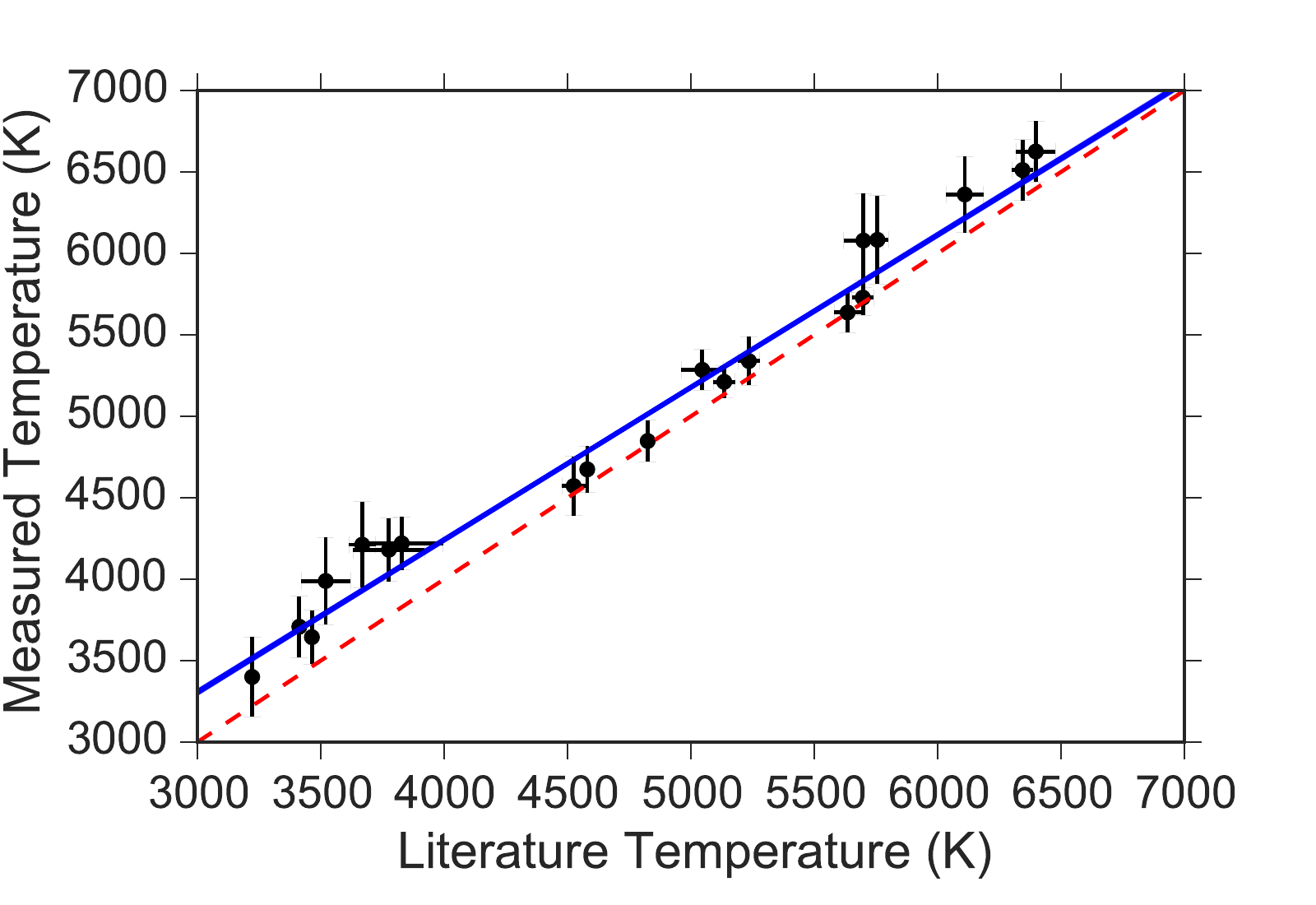}
         }
         \subfigure[IGRINS]{
             \label{fig:error:c}
             \includegraphics[scale=0.45]{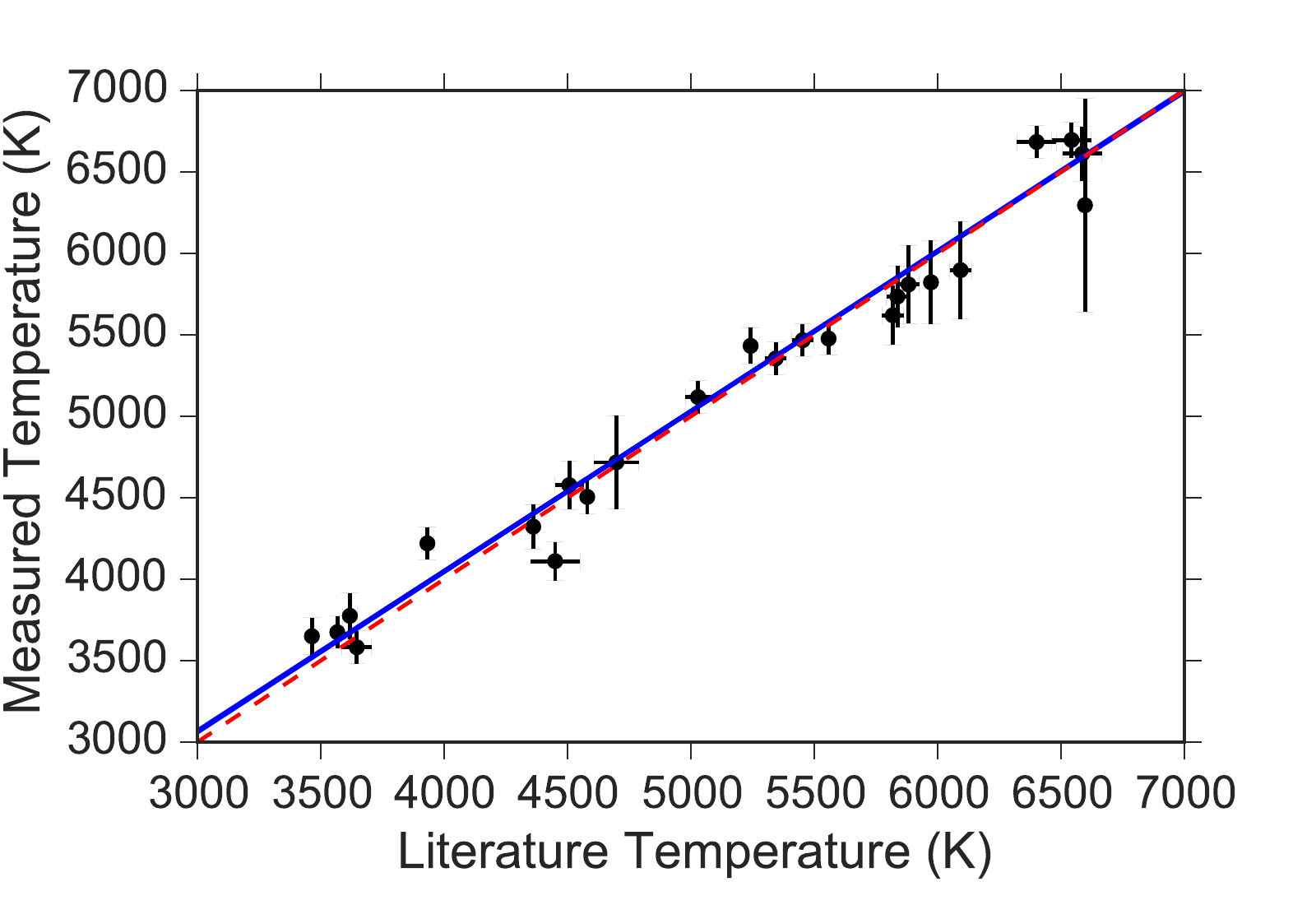}
         }
         \subfigure[CHIRON]{
             \label{fig:error:d}
             \includegraphics[scale=0.45]{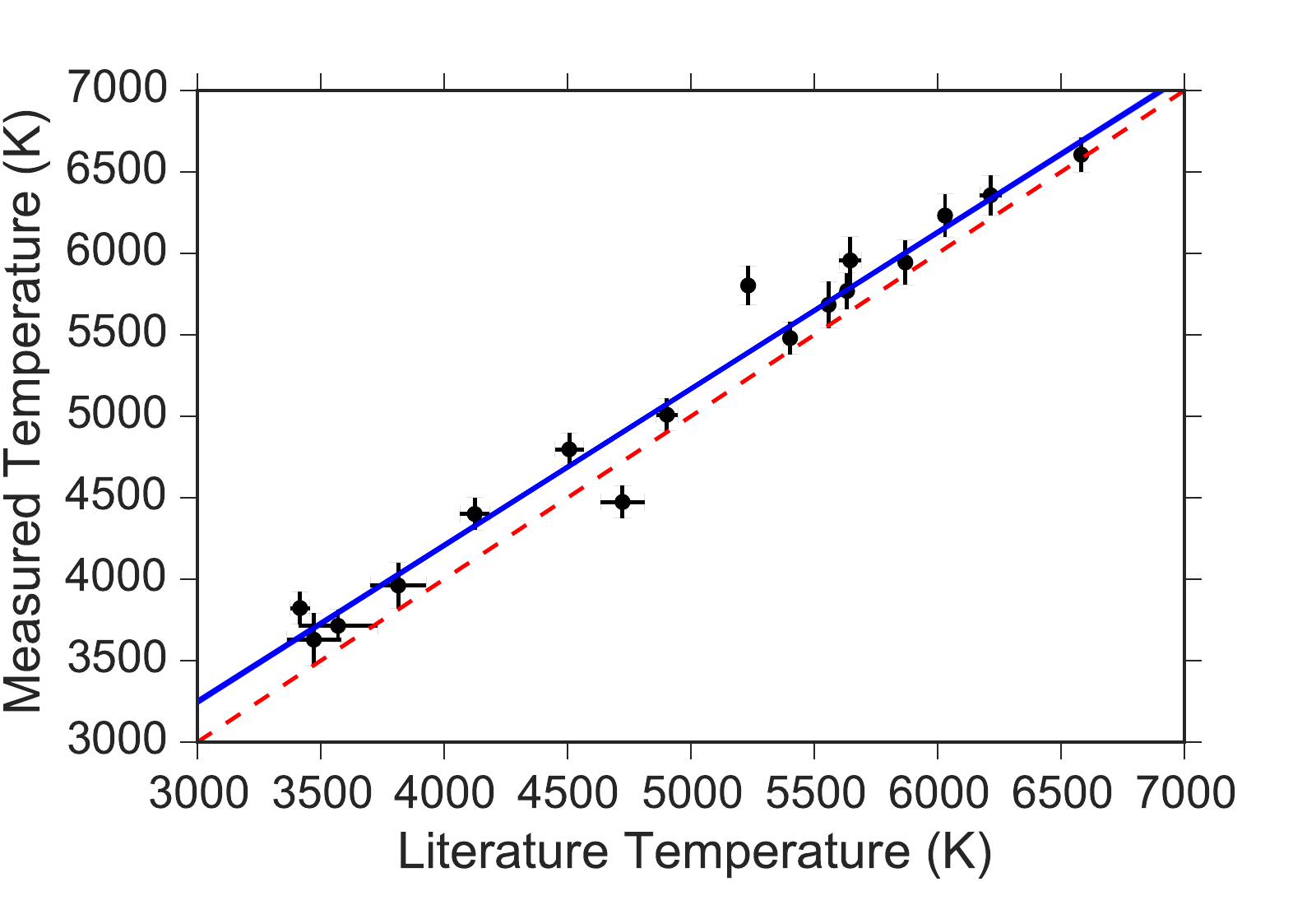}
         }
         \caption{Correspondence between the companion temperature measured with the direct spectral detection method, and the actual (literature) values. In all figures, the red dashed line has unity slope, the values with uncertainties are the measurements from the synthetic binary observations (see Section \ref{subsec:systematics}), and the blue lines are the line of best-fit through the data. There is significant bias in all of the measurements except for those using the near-infrared IGRINS instrument.}
         \label{fig:error}
\end{figure}

\section{Parameter Determination}
\label{subsec:systematics}

In the absence of noise, the CCF of an observed spectrum with a perfect model will have a value of 1 at the radial velocity of the star. As the model becomes a worse representation of the data, the peak height of the resulting CCF will decrease. Thus the CCFs act in a similar way as a $\chi^2$ map of the parameter space, allowing us to measure the effective temperature, metallicity, and rotational broadening of the secondary star. However, the presence of noise and the imperfections in the model spectra cause the measured values to deviate from the true parameters of the secondary star. 

To measure the impact of both random and systematic noise on the parameter estimation, we created several hundred synthetic binary systems for each instrument used in our program. We made the synthetic binary systems by combining the early-type star spectra from Table \ref{tab:earlycal} with those of the late-type stars in Table \ref{tab:latecal} in every possible combination, provided both observations came from the same instrument. By combining actual observations of early-type and late-type stars, our synthetic binary observations retain any instrument-specific effects that may impact the temperature estimation. We scaled the flux of the late-type star such that the flux ratio ($F_{\rm secondary}/F_{\rm primary}$) is ten times larger than the expected flux ratio for main sequence components. The artificial brightening relative to a real binary system is to ensure that the temperature estimation uncertainties are separate from the overall sensitivity of the method, which we discuss in Section \ref{subsec:sensitivity}. We estimate the main sequence flux ratio from the published temperature of the late type star (given in Table \ref{tab:latecal}) and the published spectral types of the primaries available on Simbad \citep{Simbad}, and convert to temperature and luminosity by using Table 5 of \citet{Pecaut2013}.

We analyzed each synthetic binary star system using the method described above, and measured the temperature ($T_m$) and variance ($\sigma_T^2$) as a weighted sum near the grid point with the highest CCF peak value, weighting by the peak CCF height at each temperature ($C_i$):

\begin{eqnarray}
\label{eqn:tmeas} 
T_m &=& \sum_i C_i T_i / \sum_i C_i \\
\sigma_T^2 &=& \frac{\sum_i C_i (T_i - T_m)^2}{ \sum_i C_i - \sum_i C_i^2 / \sum_i C_i}
\end{eqnarray}

Each synthetic binary observation contributes a pair of measured and actual (literature) companion temperatures, and so each late-type star in Table \ref{tab:latecal} has many independent temperature measurements made with the DSD method. To determine the correspondence between measured and actual temperature, we perform a Markov Chain Monte Carlo (MCMC) fit to a straight line using the emcee code \citep{emcee}. We plot the mean and standard deviation of the measured temperatures in Figure \ref{fig:error}, along with 300 MCMC samples for the linear fit and the line of unity slope. The MCMC samples give posterior probability distributions for the parameters $a$ and $b$ relating the actual temperature ($T_a$) to the measured temperature ($T_m$) through

\begin{equation}
T_m = a + bT_a
\end{equation}
In Section \ref{sec:results} we invert this relation to determine the actual companion temperature, given the measured temperature from Equation \ref{eqn:tmeas}.

The typical temperature uncertainty with the DSD method is $\sim 150 - 200$ K, but the optical instruments systematically overestimate the companion temperature. \emph{The error analysis is therefore not just important to measure the parameter uncertainties, but also to get the correct answer.} We suspect the systematic biases come from a mismatch between the Phoenix model spectrum template and the real spectrum of a late-type star. The biases are different for each instrument because the instruments have different wavelength ranges, and so the spectral lines that contribute most to the cross-correlation function are different.

\section{Detection Sensitivity}
\label{subsec:sensitivity}

\begin{figure}
  \centering
  \includegraphics[width=\columnwidth]{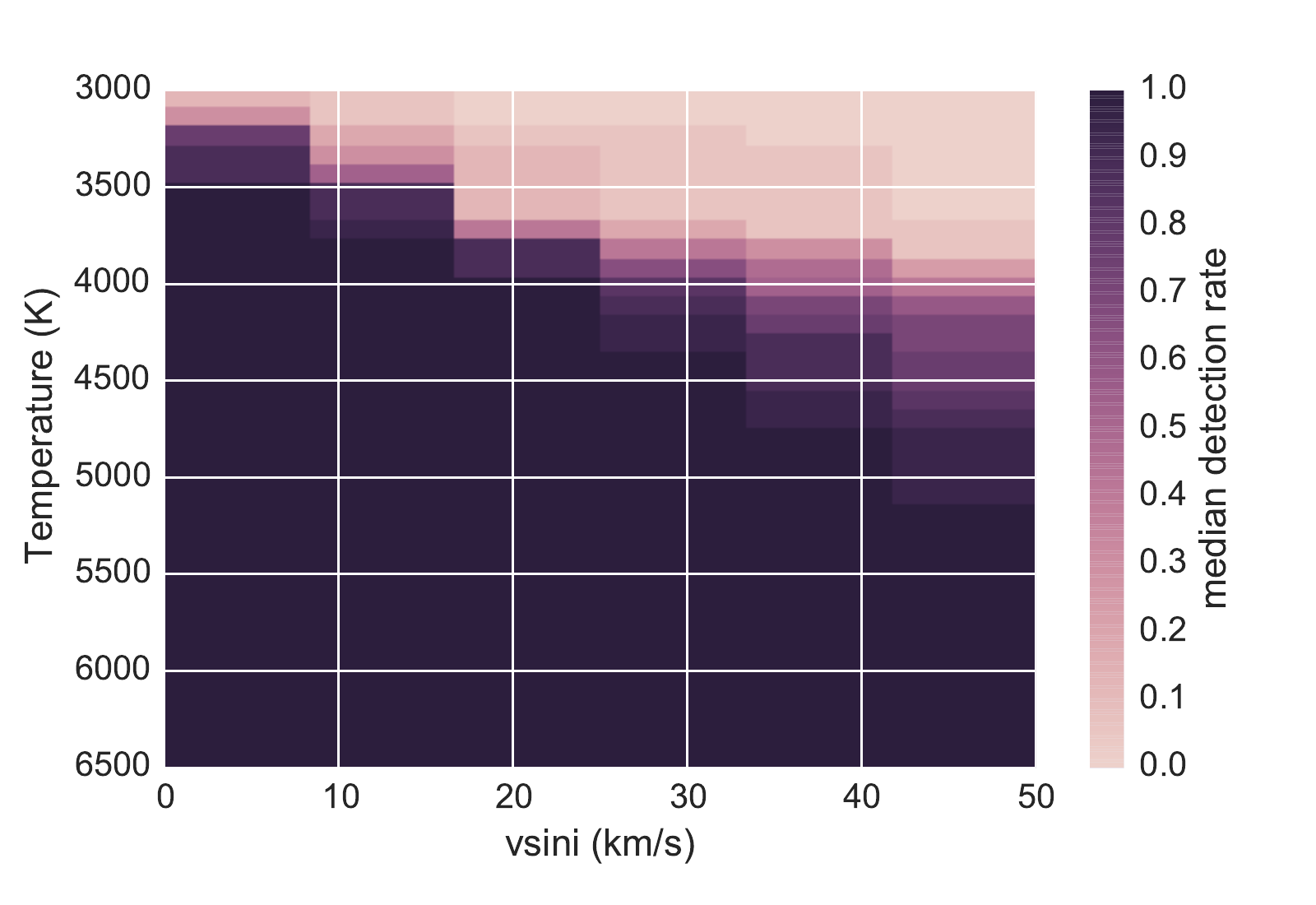}
  \caption{Median detection rate as a function of companion temperature and rotation speed. Each cell represents the median detection rate for targets with no detection in Table \ref{tab:known}. Companions represented by dark cells are detectable. See Section \ref{subsec:sensitivity} for details of the analysis.}
  \label{fig:sensitivity_2d}
\end{figure}

The detectability of a companion decreases primarily as the contrast between it and the primary star increases. Rotation plays an important role in the detection rate as well, since the cross-correlation function derives most of its power from narrow spectral features. We follow a similar strategy as above to estimate the detection rate as a function of temperature and rotational velocity for each star, with the key differences that we scale the model spectra to replicate a binary star observation with main sequence observations (rather than scaling the companion to ten times main sequence), and that we add \emph{Phoenix model spectra} for late-type stars to the data instead of real spectra. We use synthetic spectra so that we can use a finer grid of temperatures and rotational broadening and not be limited by the temperatures or the temperature estimation uncertainties of real late-type stars. However, since we are comparing models to models any mismatch between the model spectrum and the real spectrum of a star of that temperature will tend to make the sensitivity calculations somewhat optimistic. This will have the largest impact for very cool stars, where the difficult to model molecular absorption is more important.

For each observed early-type star, we generate several synthetic binary star observations by adding model spectra for stars with $\rm T_{eff} = 3000 - 7000$ K in steps of 100 K and rotational velocities $v\sin{i} = 0 - 50$ $\rm km \cdot s^{-1}$ in steps of 10 $\rm km \cdot s^{-1}$. For each temperature and $v\sin{i}$ combination, we make 17 independent synthetic observations by adding the model to the data with a radial velocity shift between -400 to 400 $\rm km \ s^{-1}$ in steps of 50 $\rm km \ s^{-1}$. We label a companion as detected if the highest peak in the CCF of the synthetic data with the model spectrum of the same temperature is within 5 $\rm km \ s^{-1}$ (the approximate instrumental broadening) of the correct velocity. 

The median detection rate for targets in Table \ref{tab:known} is shown in Figure \ref{fig:sensitivity_2d}\footnote{A file with the results of the sensitivity analysis, as well as sensitivity figures similar to Figure \ref{fig:sensitivity_2d} for each individual target, are available at this url: \url{https://github.com/kgullikson88/DSD-Paper}}. We can usually detect very cool stars if they are slowly rotating, but the sensitivity quickly degrades as the companion $v\sin{i}$ increases. Cool stars spin down as they age \citep{Barnes2003} so the rotation speed dependence is equivalent to an age dependence. We estimate the impact of rotation on our detection method by using the gyrochronology relation given in \cite{Barnes2010b}: 

\begin{equation}
\frac{k_Ct}{\tau} = \ln\left ( \frac{P}{P_0} \right ) \frac{k_Ik_C}{2\tau^2} (P^2 - P_0^2)
\label{eqn:gyro}
\end{equation}

In Equation \ref{eqn:gyro}, $k_C$ and $k_I$ are constants fit to data with known ages and rotation periods, $P$ and $P_0$ are respectively the current and zero-age main sequence (ZAMS) rotation periods, $\tau$ is the convective turnover time scale and $t$ is the current age of the star. We use the same values that \cite{Barnes2010b} use for the constants:

\begin{itemize}
\item $k_C = 0.646$ day/Myr
\item $k_I = 452$ Myr/day
\end{itemize}

\begin{figure}
    \centering
    \includegraphics[width=80mm]{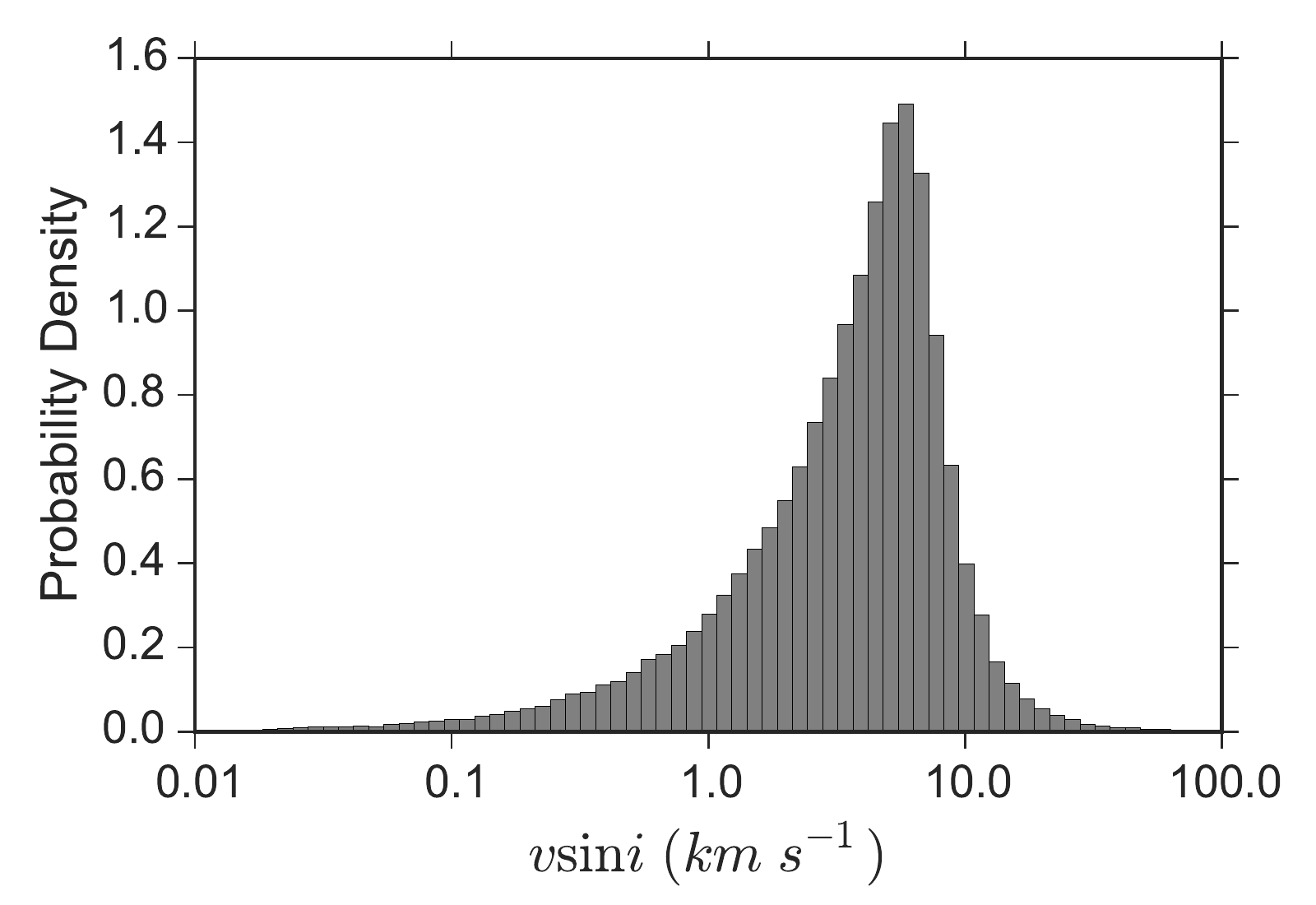}
    \caption{Typical probability density function for companion rotational velocity $v\sin{i}$. The distribution peaks near $\sim 5-10$ $\rm km \ s^{-1}$ and extends to very high velocities. Note that the x-axis is log-spaced to more clearly show the tails of the distribution.}
    \label{fig:vsini}
\end{figure}

We use Equation \ref{eqn:gyro} to estimate the expected rotation period for a companion star of given temperature and age as follows: First, we convert from temperature to convective timescale ($\tau$) by interpolating Table 1 in \citet{Barnes2010a}. Next we sample an appropriate probability density function (PDF) for the age of the binary system; if the primary star was analyzed in \citet{David2015}, we use their posterior age PDFs. Otherwise, we use a uniform PDF from the Zero Age Main Sequence (ZAMS) age of the primary star to its main sequence lifetime (typically 10-200 Myr for our sample). Following the discussion in \cite{Barnes2010b}, we uniformly sample initial rotation periods from 0.2 - 5 days for all stars. We estimate the current rotation period for each pair of age and initial rotation period samples using Equation \ref{eqn:gyro} to build up a PDF of current rotation periods. We transform the period distribution into a PDF for $v\sin{i}$ using the main sequence radius of a star of the given temperature, obtained by interpolating Table 1 of \cite{Barnes2010a}, and a uniform sampling of inclinations ($\sin{i}$). Figure \ref{fig:vsini} shows a typical $v\sin{i}$ distribution, which peaks near $\sim 5-10$ $\rm km \ s^{-1}$ and has a long tail extending to $\sim 40-50 \ \rm km s^{-1}$.

\begin{figure}
        \centering
        \includegraphics[width=80mm]{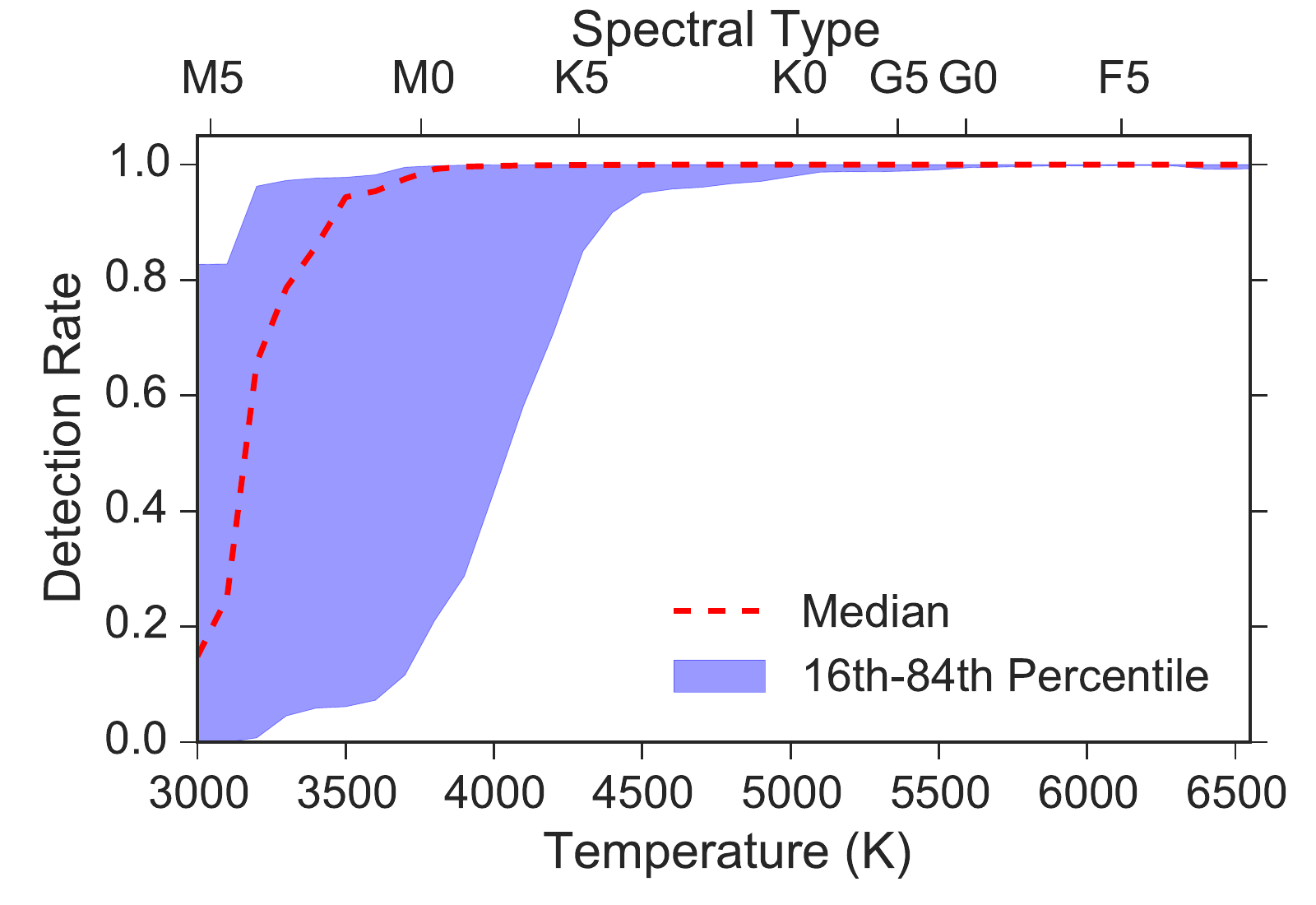}

         \caption{Summary of the detection rate as a function of temperature for the sample stars (Table \ref{tab:known}) in which we do not detect a companion. The red dashed line gives the median detection rate, and the blue filled area illustrates the range across different primary stars. The direct spectral detection method can detect companions as late as M0 for most of our targets.}
         \label{fig:sensitivity}
\end{figure}

By combining the sensitivity calculations described above with the $v\sin{i}$ samples, we marginalize over the expected rotation periods of the secondary stars to get simpler curves of detection rate as a function of companion star temperature. We show the median and approximate range of the marginalized detection rate in Figure \ref{fig:sensitivity}. The DSD method can reliably detect companions as cool as 3700 K in most cases, although the primary star spectral type plays a dominant role in setting the coolest detectable companion. 

Companions with $T \gtrsim 6250$ K, the canonical limit at which the convective zone is too small to transfer angular momentum to the stellar wind and spin down the star \citep{Pinsonneault2001}, may have rotational velocities comparable to that of the primary star. In that case, estimating the primary star spectrum with a gaussian filter may remove much or all of the companion spectrum. Since these are the types of stars with less extreme flux- and mass-ratios, they are easier to detect with more conventional methods. However, this shortcoming could be overcome by using model spectra for the primary star as in \citet{Kolbl2015}. In this work, we have optimized the method for finding cool companions.

\begin{figure*}
  \centering
  \subfigure[]{
       \label{fig:detections:a}
       \includegraphics[width= 75mm]{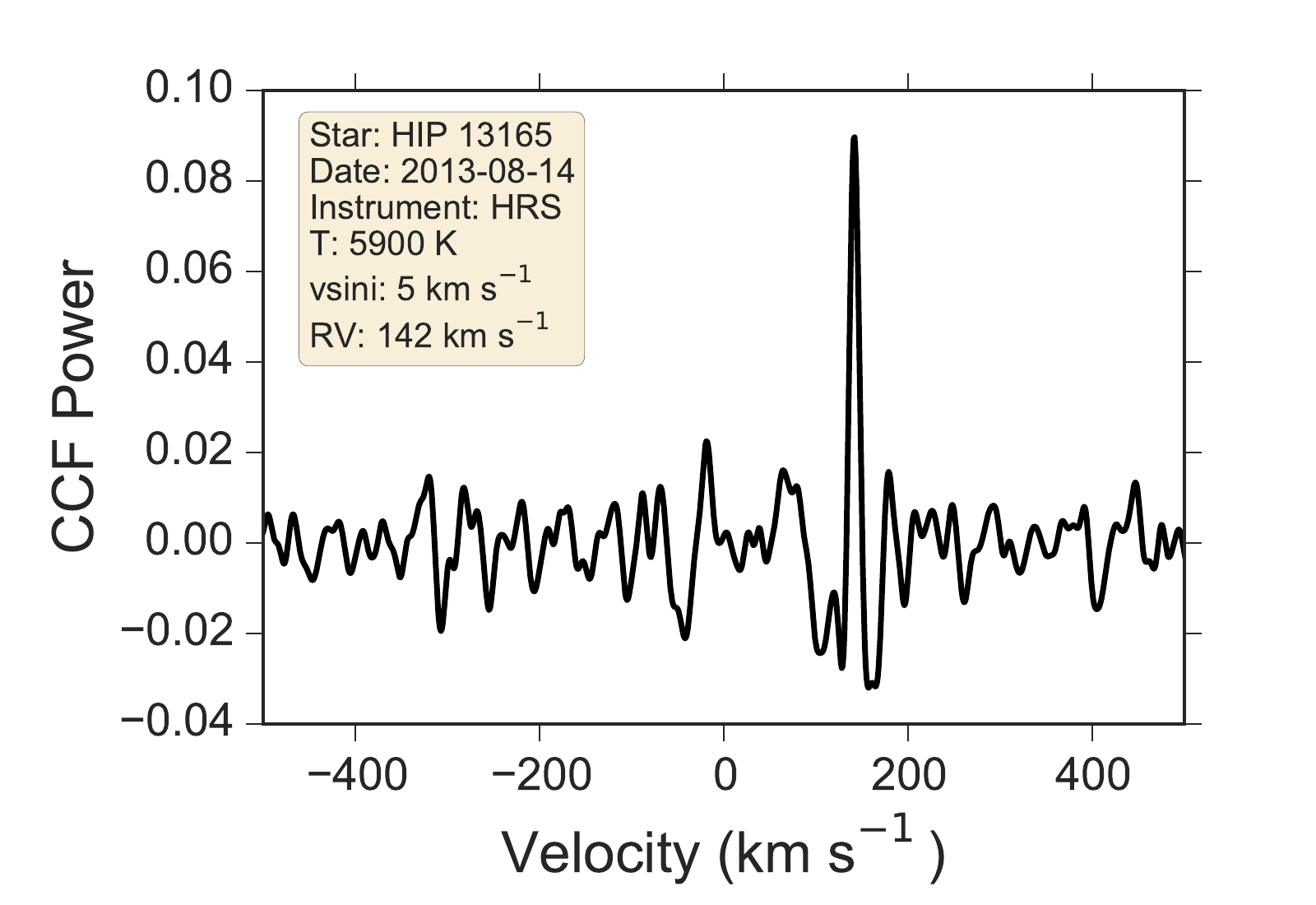}
  } \quad
  \subfigure[]{
       \label{fig:detections:b}
       \includegraphics[width= 75mm]{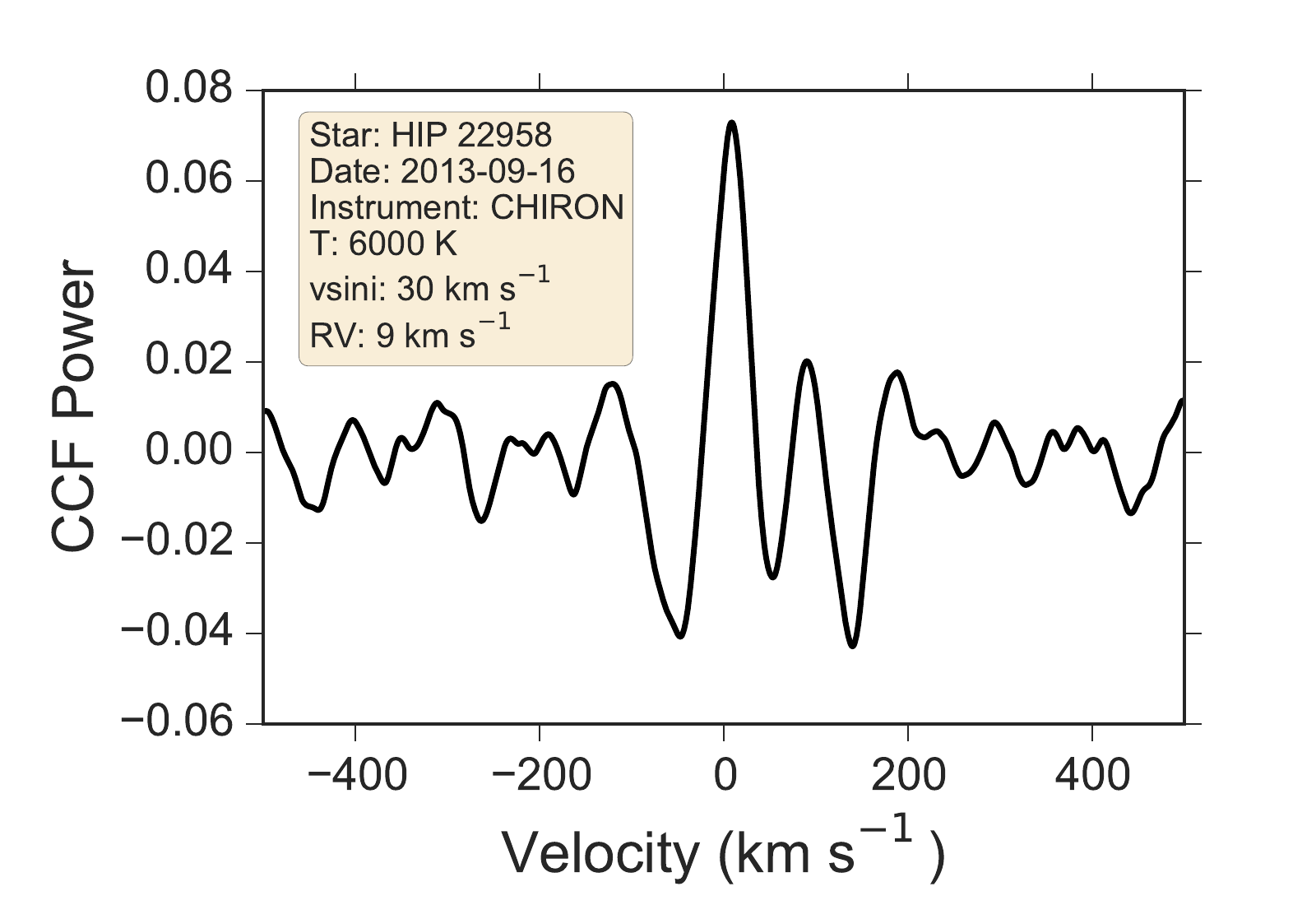}
  }
  
  \subfigure[]{
       \label{fig:detections:c}
       \includegraphics[width= 75mm]{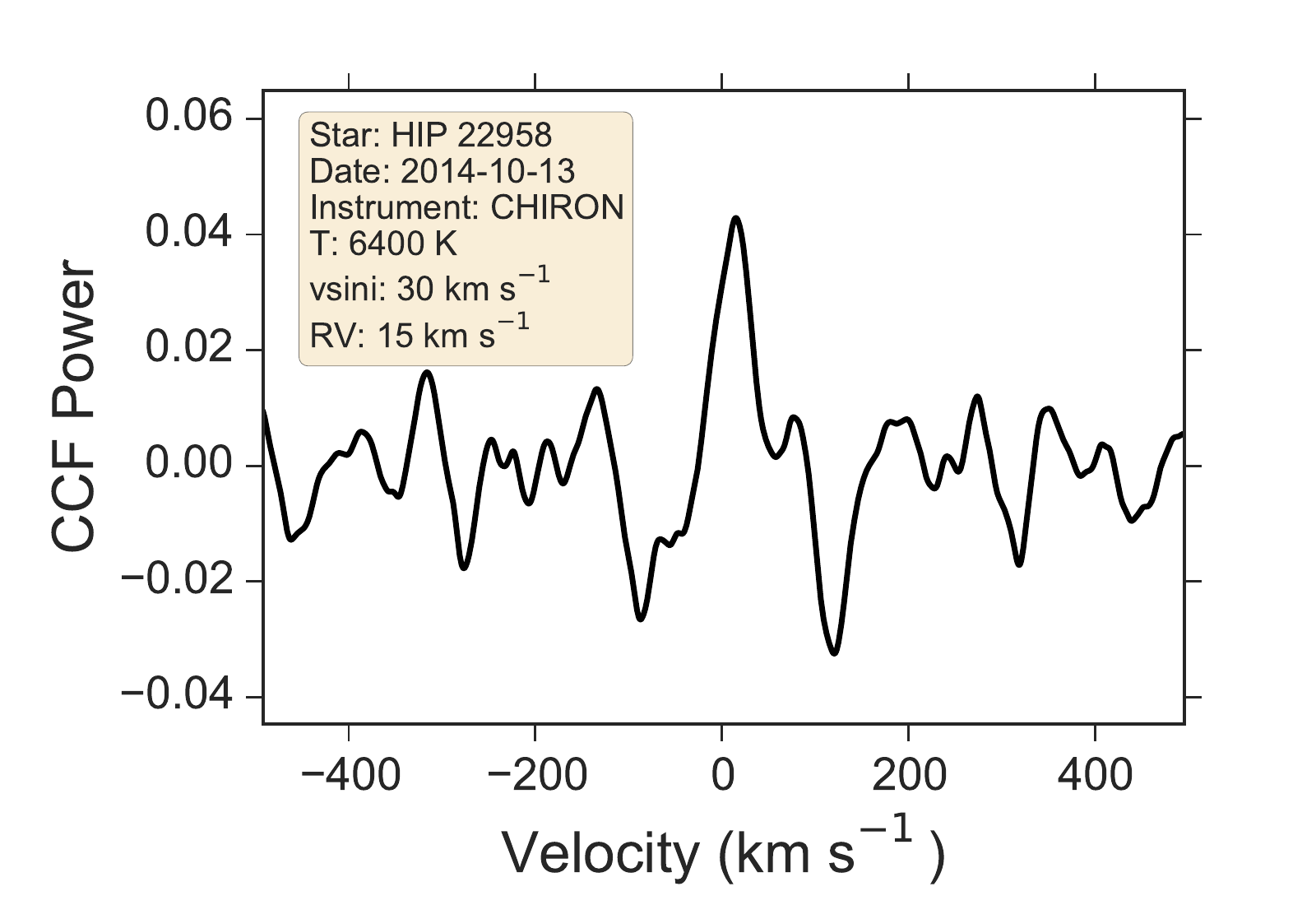}
  } \quad
  \subfigure[]{
       \label{fig:detections:d}
       \includegraphics[width= 75mm]{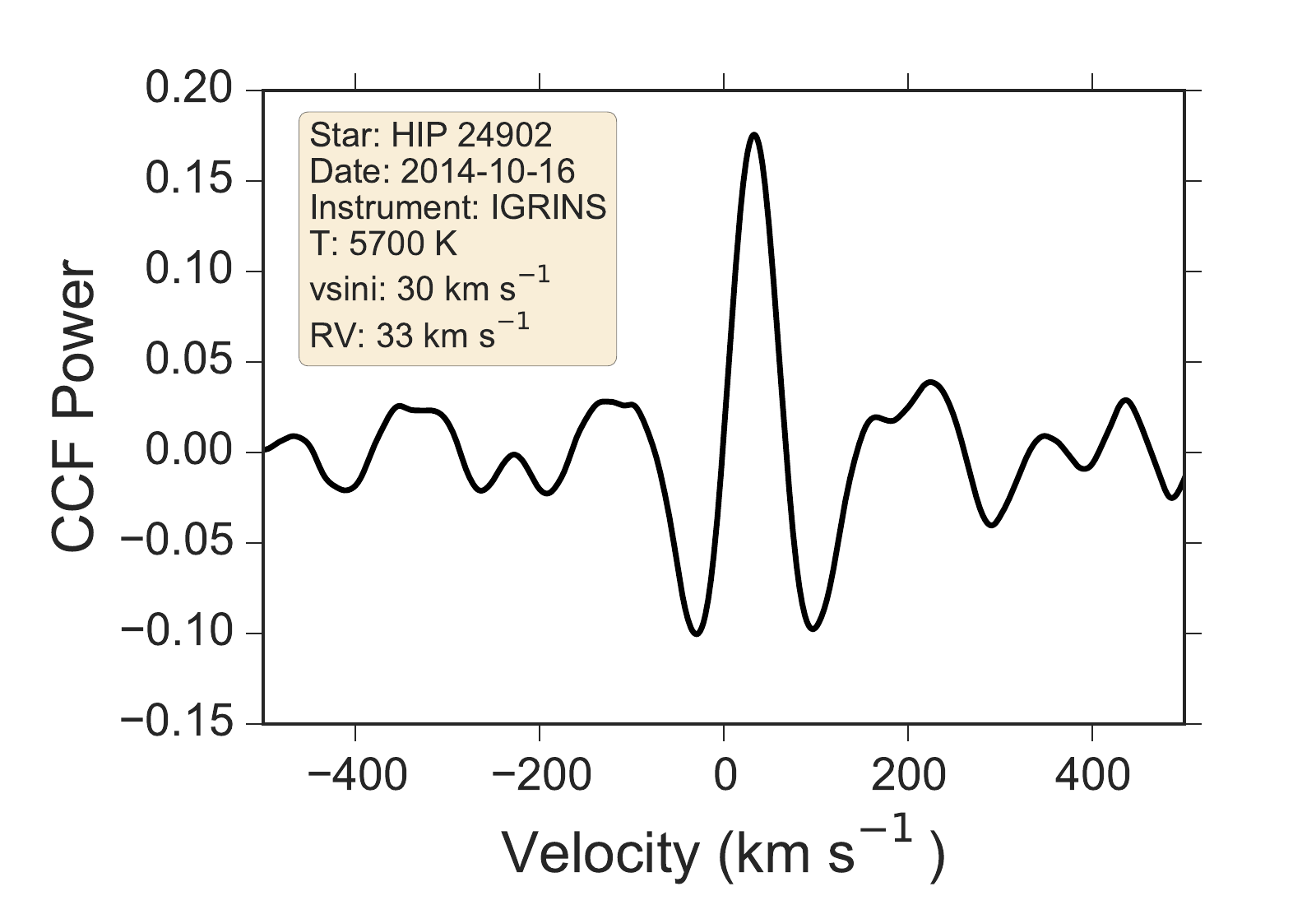}
  }
  
  \subfigure[]{
       \label{fig:detections:e}
        \includegraphics[width= 75mm]{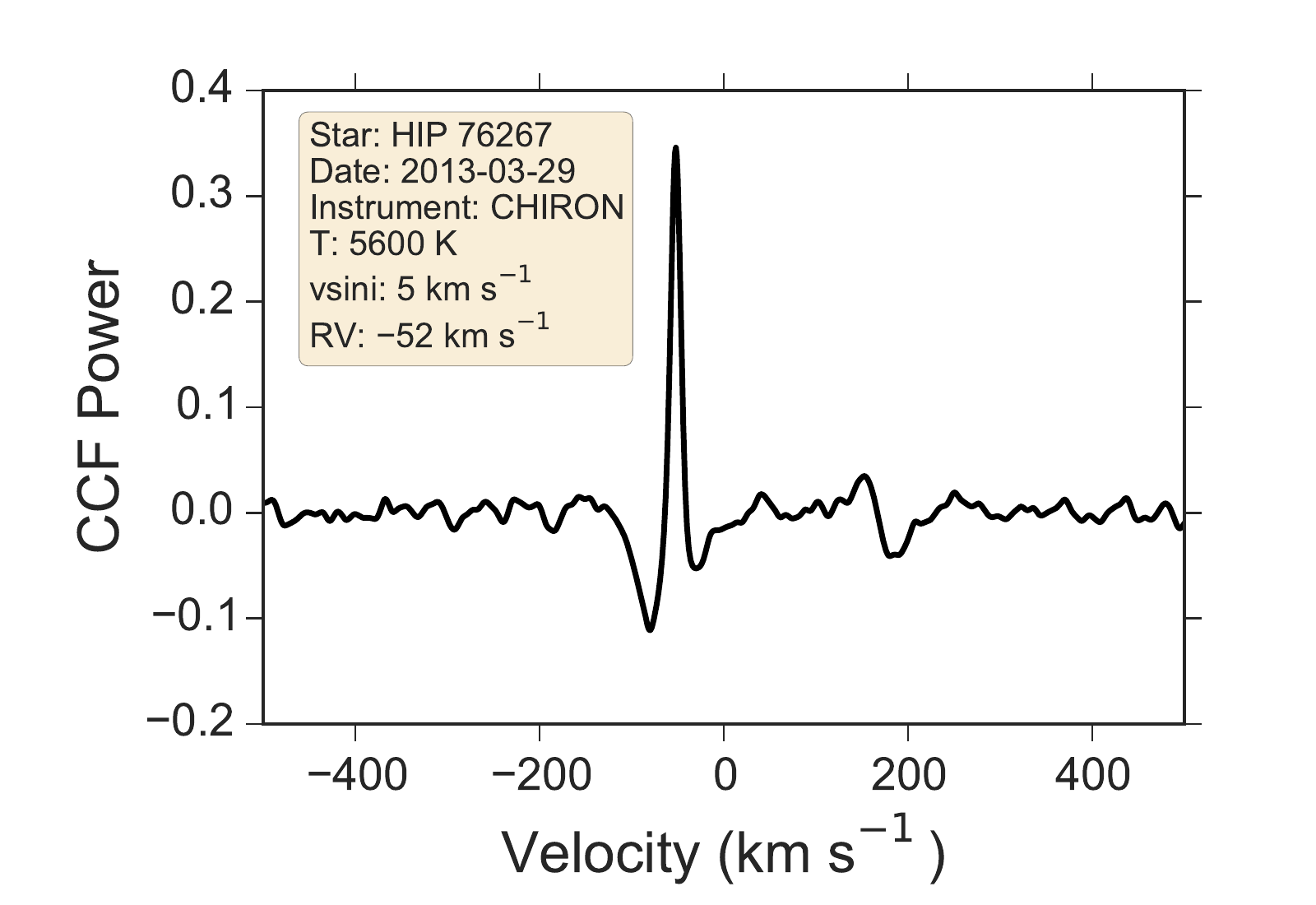}
  }
  
  \caption{Cross-correlation functions for detected companions}
  \label{fig:detections}
\end{figure*}

\begin{figure*}
 \centering
 
  \subfigure[]{
       \label{fig:detections2:a}
       \includegraphics[width= 75mm]{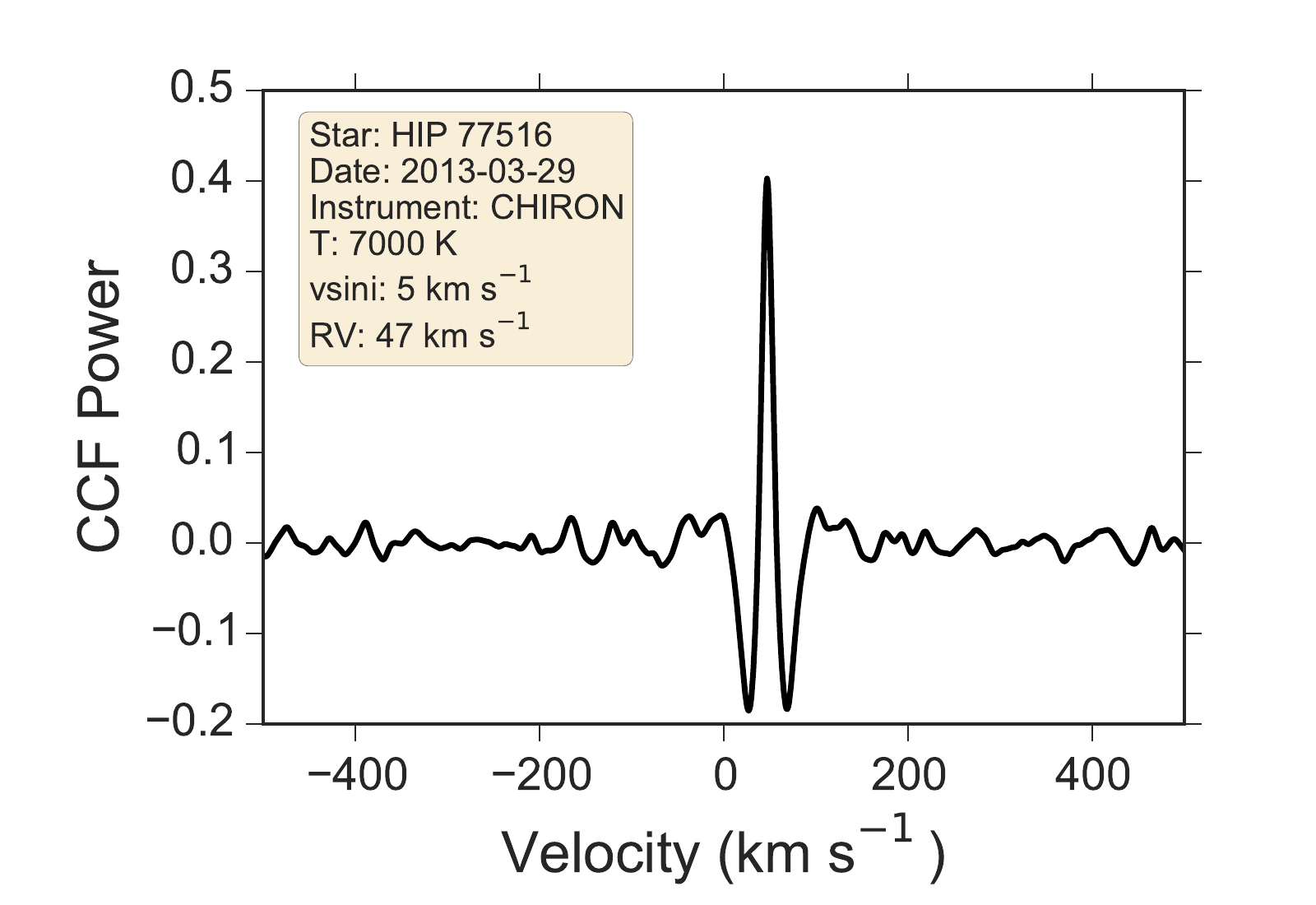}
  } \quad
   \subfigure[]{
       \label{fig:detections2:b}
       \includegraphics[width= 75mm]{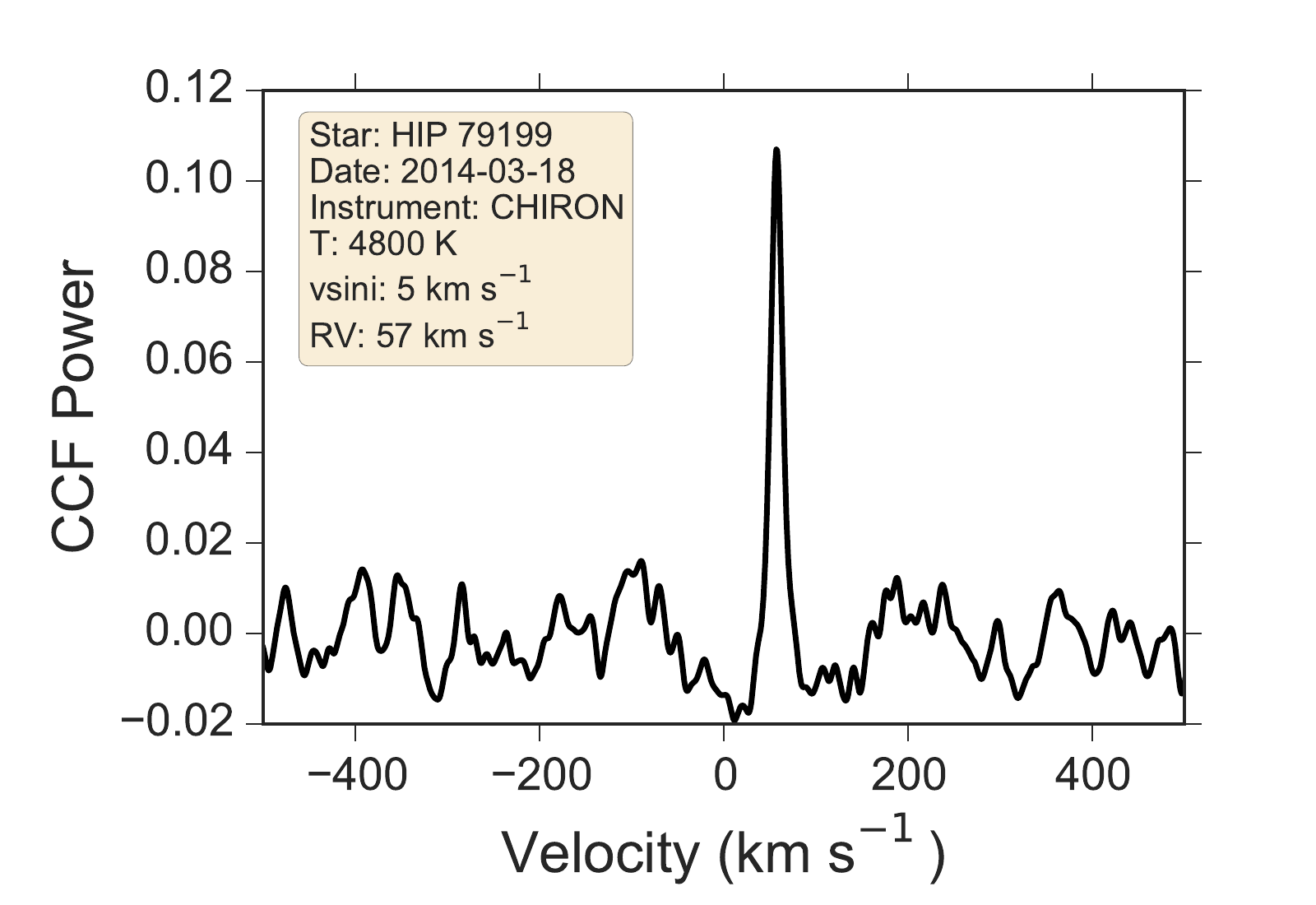}
  }
   
  \subfigure[]{
       \label{fig:detections2:c}
       \includegraphics[width= 75mm]{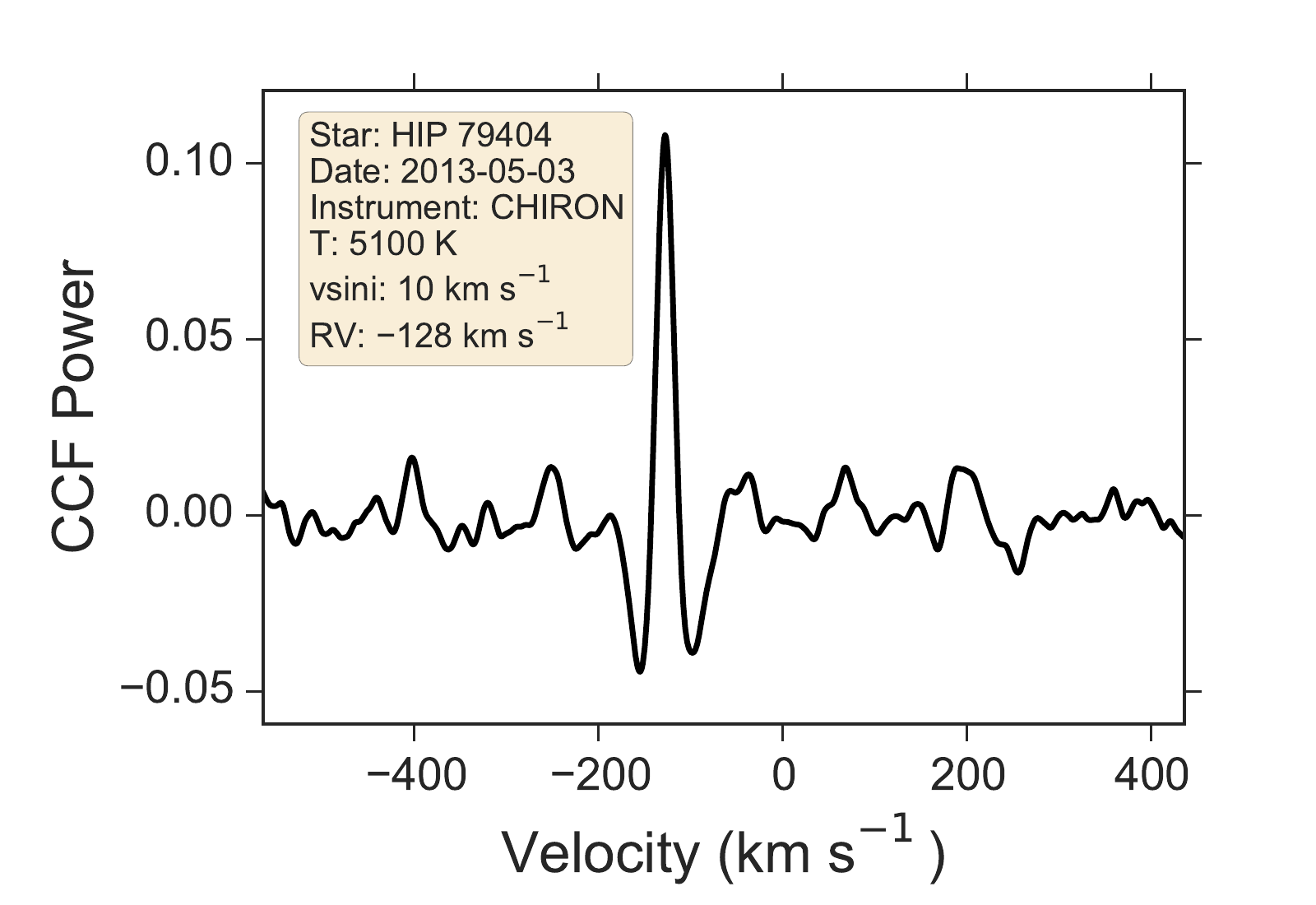}
  } \quad
   \subfigure[]{
       \label{fig:detections2:d}
       \includegraphics[width= 75mm]{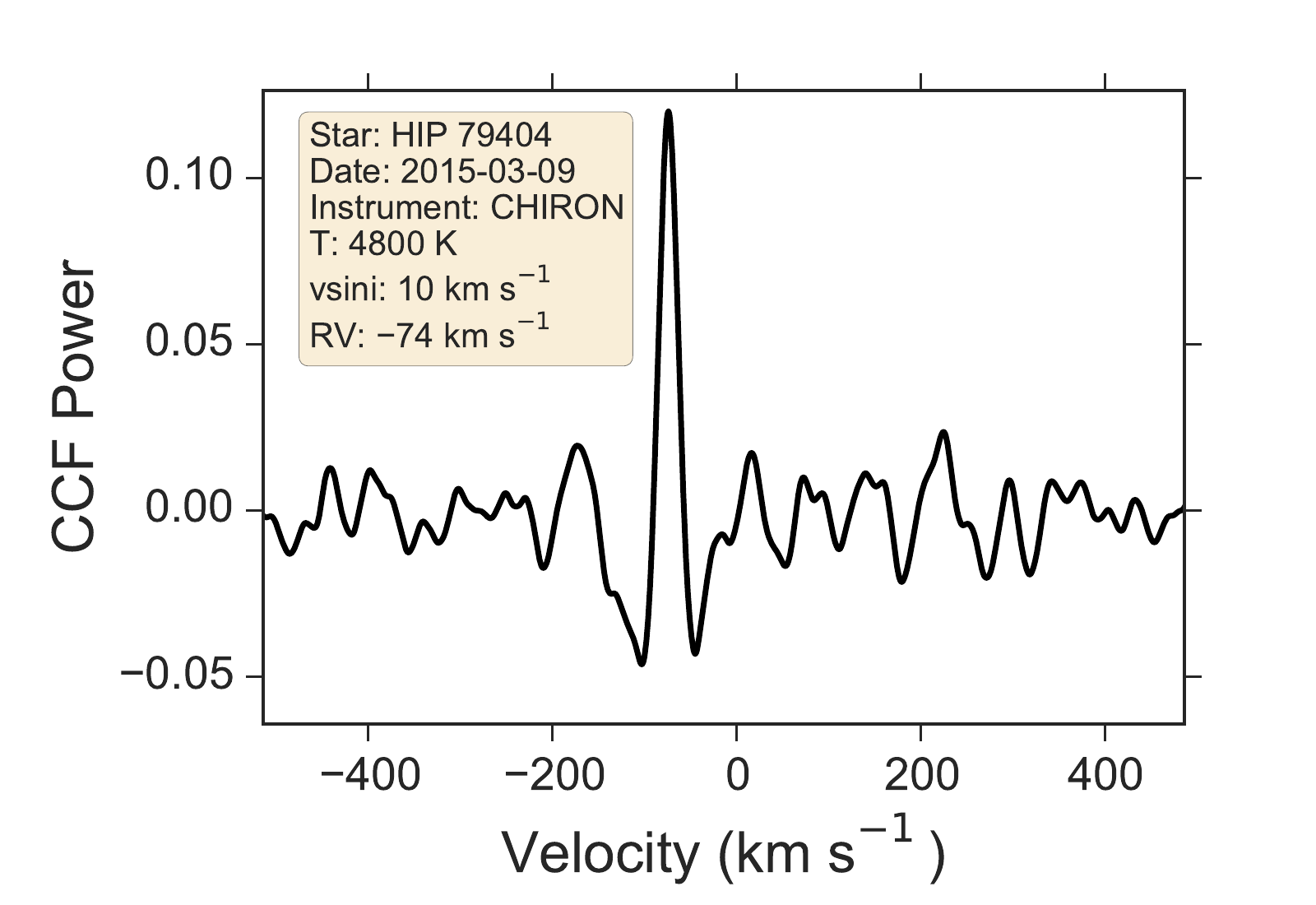}
  }
  
  \subfigure[]{
       \label{fig:detections2:e}
       \includegraphics[width= 75mm]{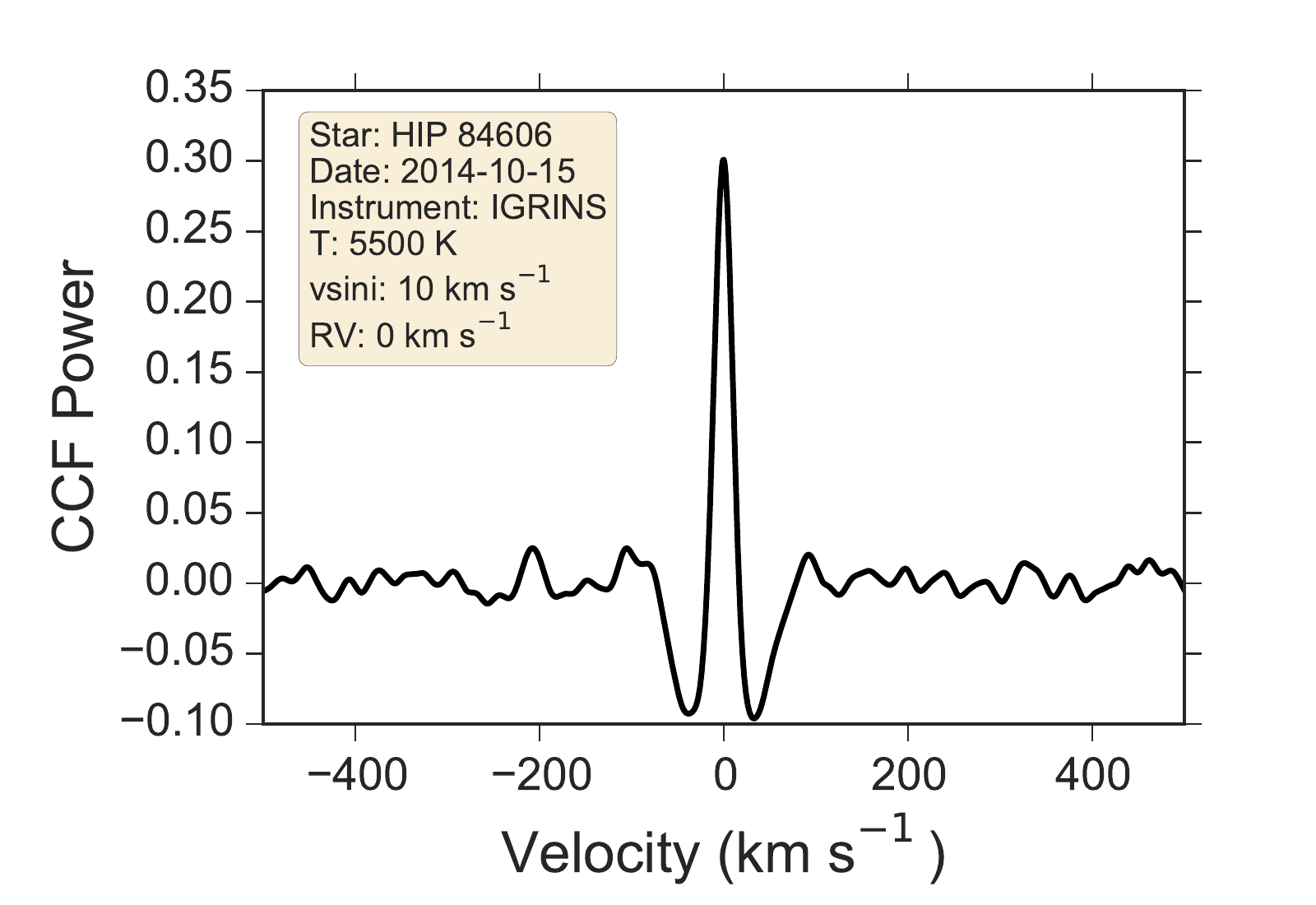}
  } \quad  
  \subfigure[]{
       \label{fig:detections2:f}
       \includegraphics[width= 75mm]{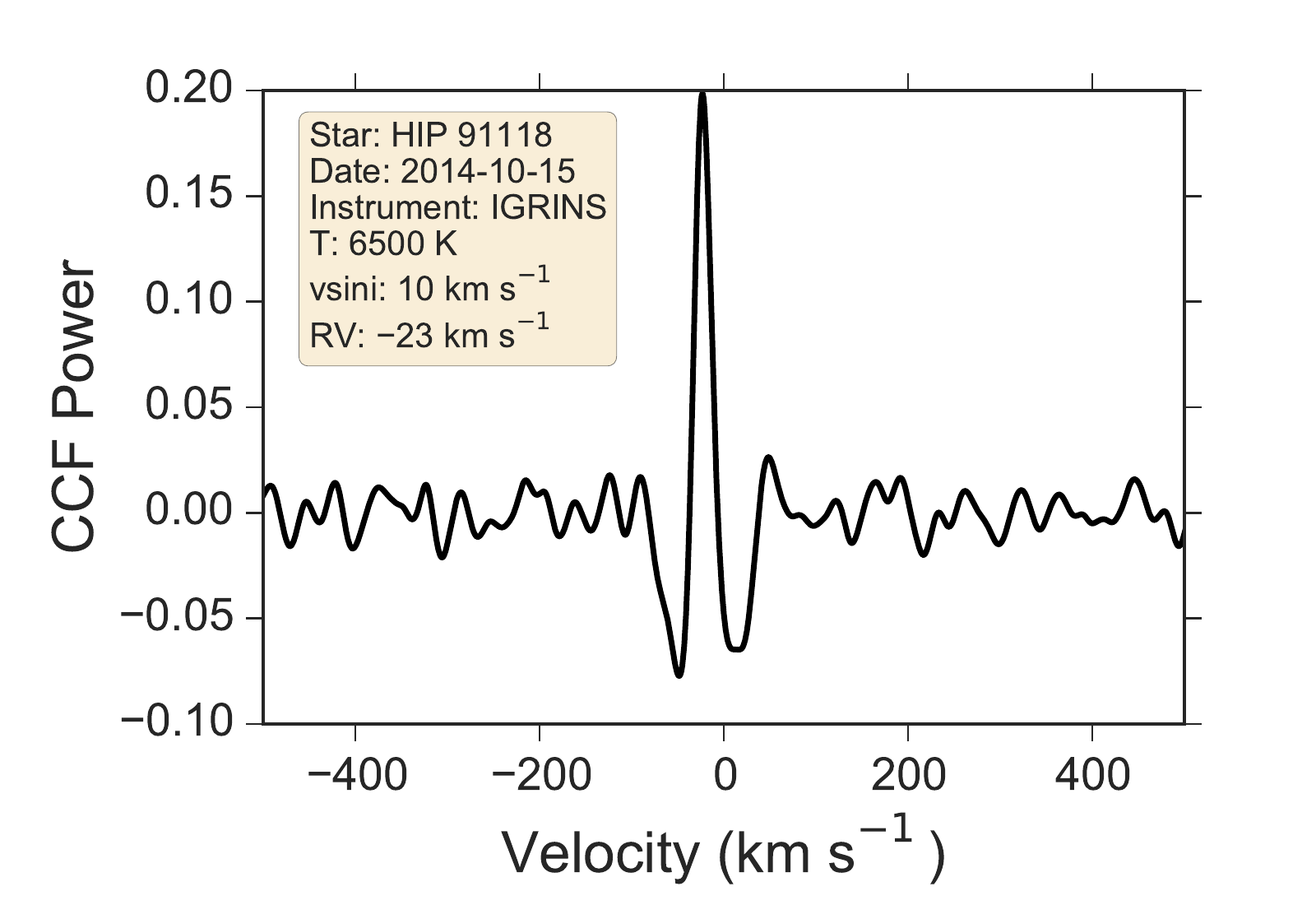}
  } 
  
  \caption{Cross-correlation functions for detected companions}
  \label{fig:detections2}
\end{figure*}

\section{Application to Known Binary Systems}
\label{sec:results}

We now use the DSD method to measure the temperatures of several known binary systems (Table \ref{tab:known}). We cross-correlate the spectra against the full grid of model spectra enumerated in Section \ref{sec:method}, and find the temperature of the companion using Equation \ref{eqn:tmeas}. We then convert the measured temperature to PDFs of the true companion temperature using the MCMC chains developed in Section \ref{subsec:systematics} (see also Figure \ref{fig:error}). For stars with multiple observations, we multiply the PDFs from each detection. Finally, we calculate the companion temperature and confidence interval from the integral of the PDF:

\begin{equation}
f = \int_{-\infty}^x {pdf(T)dT}
\end{equation}
We use as the central value the value of x such that $f=0.5$ (the median). Likewise, we calculate the $1 \sigma$ lower and upper bounds such that $f = 0.16$ and $f = 0.84$, respectively. The CCFs for the companions that we detect are shown in Figures \ref{fig:detections} and \ref{fig:detections2}. For each star, we show the CCF which has the maximum peak value and annotate the figures with the parameters. Most of the CCFs have very strong peaks. The exception is HIP 22958; however, the detection is strengthened by the fact that we observed this star twice and measured a similar temperature both times. The CCFs for HIP 22958 and HIP 24902 demonstrate the adverse effect a large companion rotational velocity has on the detection significance.

\subsection{Comparison to Literature Data}
\label{subsec:expected_teffs}
We use the literature data to predict an expected temperature for each companion in order to directly compare our measurements to previous results. The procedure outlined in Section \ref{sec:obs} using the magnitude difference or orbital information alone produces reasonable estimates, but in many cases there is additional information in the literature to refine the estimates. The refined estimates are described below.

HIP 76267 and HIP 84606 are found in the \cite{David2015} sample; we use the mass and temperature estimates provided there rather than going through the Simbad spectral type and assuming main-sequence relationships.

\citet{Shatsky2002} provide a color estimate of the companion star to HIP 79199 ($J-K = 0.57 \pm 0.12$). We convert this directly into a temperature estimate through Table 5 of \citet{Pecaut2013}.

\citet{Zorec2012} find fundamental parameters for HIP 22958, and determine a temperature slightly cooler and luminosity much greater than the spectral type (B6V) would suggest. Because of this the usual analysis, which uses main sequence relationships, results in a biased answer. We estimate the companion temperature by assuming that the companion \emph{does} follow the main sequence relationships as described in Section \ref{sec:obs}, but sample the uncertainty distributions given in \citet{Zorec2012} for the temperature and radius of the primary star.

\begin{figure}
        \centering
        \includegraphics[width=80mm]{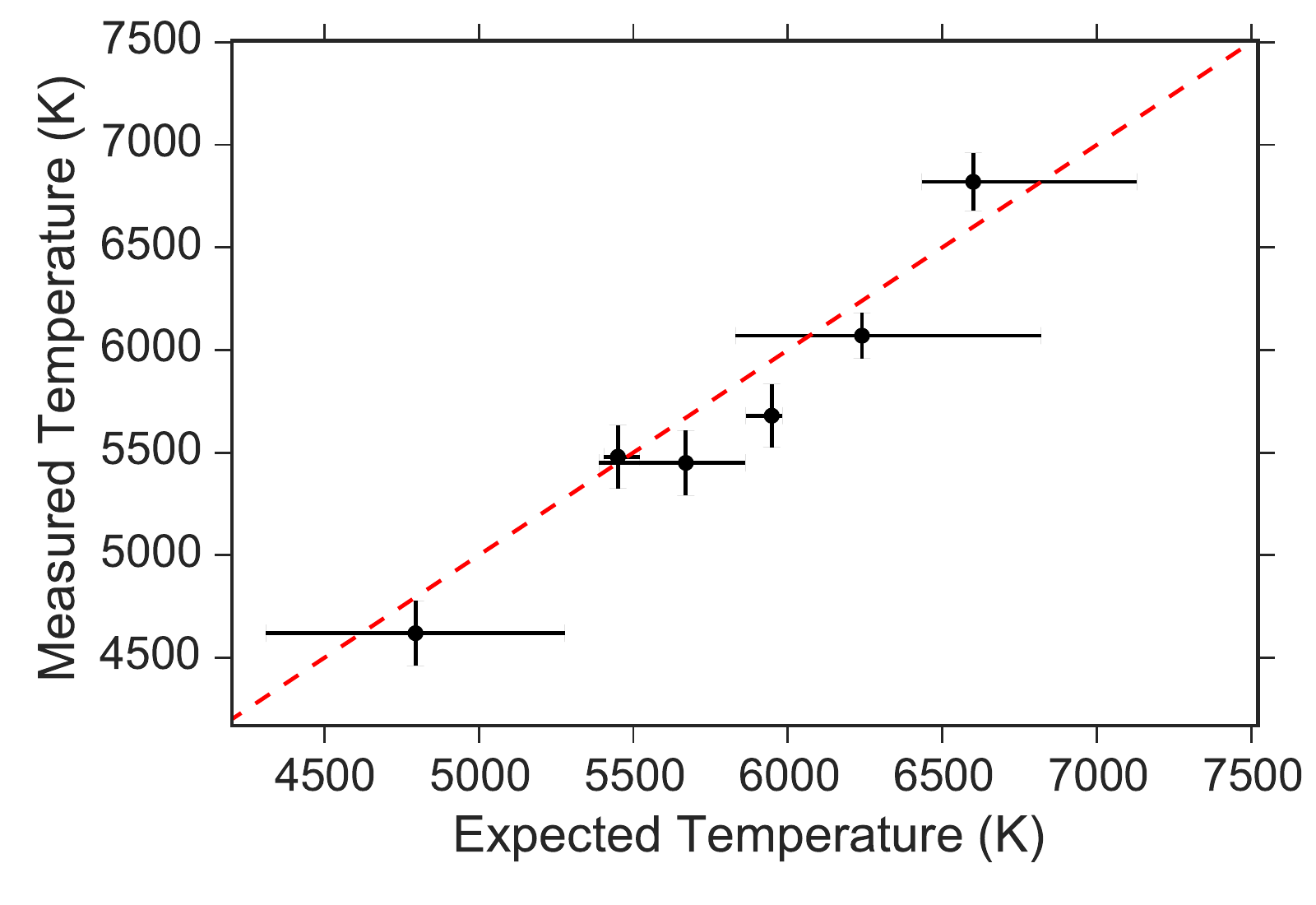}
        
        \caption{Temperature comparison for binaries with known secondary spectral types. The x-axis shows the companion temperature expected from the literature data (see Section \ref{subsec:expected_teffs}).}
         \label{fig:known}
\end{figure}

We compare our companion temperature measurements from the DSD method to the estimates described above in Figure \ref{fig:known}. There is overall excellent agreement between the temperatures, with 5/6 falling within $1 \sigma$ of equality. We test for a bias ($\Delta$) between the measured temperatures ($T_m$) and expected temperatures ($T_a$) with the equations

\begin{eqnarray}
\Delta &= \sum_i(T_{m,i} - T_{a,i}) \\
\sigma_{\Delta}^2 &= \sum_i (\sigma_{T_{m,i}}^2 + \sigma_{T_{a,i}}^2)
\end{eqnarray}
which results in $\Delta = -580 \pm 770$ K. Our temperature measurements are consistent with the expected temperatures.

We list our measurements as well as the expected temperatures described above in Table \ref{tab:measured}. The expected $v\sin{i}$ values come from application of Equation \ref{eqn:gyro} as described in Section \ref{subsec:sensitivity}. While we do give the measured $v\sin{i}$ and metallicity for our detections, the accuracy of these parameters is not calibrated and is determined with a coarse grid; the values should only be taken as rough estimates. We do note that most of the measurements have $[Fe/H] = -0.5$. This is likely a measurement bias since we do not expect the binary systems to have significantly sub-solar metallicity. As metallicity increases, so do the line depths of most of the lines in the spectrum. Any lines that are poorly modeled will then have a larger negative impact on the resulting CCF; thus the bias towards low metallicity is likely a result of imperfect model atmosphere templates. We do not attempt to identify the poorly modeled lines in this work.

\subsection{Non-detections}
\label{sec:nondetections}
There are many companions in Table \ref{tab:known} that we do not detect. Most of these are single-lined spectroscopic binaries (Table \ref{tab:specdata}), and are likely too cool to detect with our data; very high signal-to-noise spectra with a near-infrared instrument such as IGRINS may uncover them. Several of the remaining un-detected companions have expected temperatures $T  > 6250$ K, and so are likely to be rapid rotators. Since the cross-correlation function gets most of its power from  sharp spectral features, these rapidly rotating companions are difficult to detect (see Figure \ref{fig:sensitivity_2d}).  

\begin{figure}
        \centering
        \includegraphics[width= 80mm]{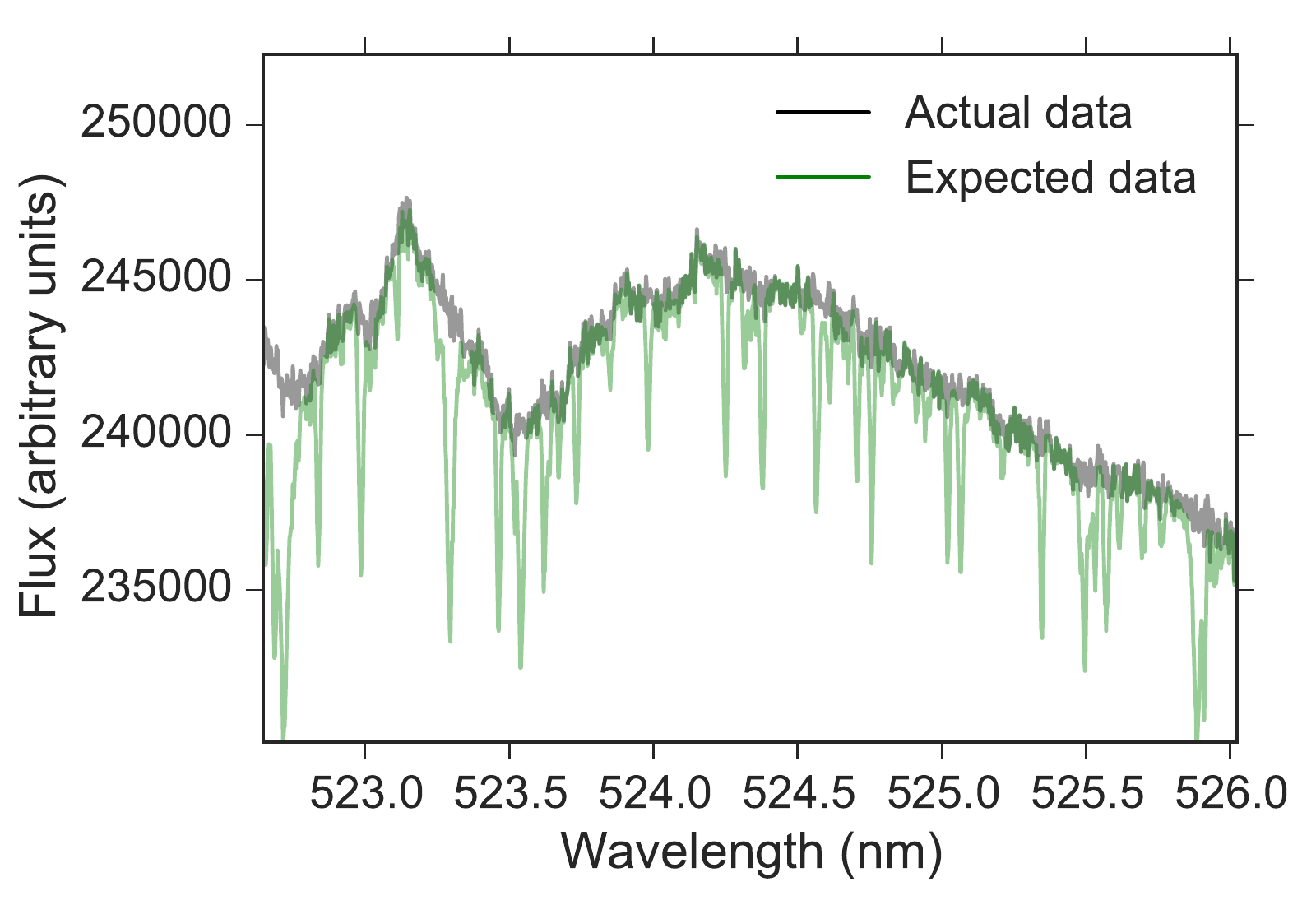}
         \caption{Observed (black) and expected (green) spectra for the known binary system HIP 88290. At the expected flux ratio, the spectral lines from the companion should be easily visible. }
         \label{fig:expected}
\end{figure}

Finally, HIP 88290 is hot enough and expected to be rotating slowly enough that we should be able to easily detect it. In fact, we would expect to be able to directly see the companion in the spectra (the green lines in Figure \ref{fig:expected}). The fact that we do not see the composite spectrum or see a peak in the cross-correlation function implies that the companion must be rotating with $v\sin{i} > 50 \rm km s^{-1}$, much more quickly than Equation \ref{eqn:gyro} predicts, that the primary is a giant and therefore much brighter than main-sequence relationships suggest, or that the companion fell outside the spectrograph slit. This star is in the \cite{David2015} sample and has an effective temperature and mass consistent with main sequence, so we can rule out the giant primary possibility. Additionally, the binary separation is $0.47''$ \citep{Tokovinin2015} and the CHIRON spectrograph has a $\sim 2.7''$ diameter fiber; light from the companion is guaranteed to fall on the slit.

\section{Discussion and Conclusions}
\label{sec:conclusions}
We have presented and extensively characterized the direct spectral detection method for finding companions to intermediate-mass stars using high-resolution cross-dispersed \'echelle spectroscopy. Using a very large number of synthetic but realistic binary star observations, we constrained the uncertainty and systematic errors present in determining the companion temperature with the direct spectral detection method. The typical uncertainties are of the order of 200 K across all instruments used in this study, with a systematic offset of similar magnitude (for the optical instruments). We used the synthetic binary star analysis to calibrate the direct spectral detection method for the four instruments used in this study between the temperatures $ 3500\ K < T < 6500\ K$.

We also estimated the sensitivity to detection of companions with a range of temperature and $v\sin{i}$ by creating a second set of synthetic companions. The method can detect companions as late as M0 in most cases, although the lower limit depends on the primary star spectral type, the signal-to-noise ratio achieved, and the instrument used. The median detection limit corresponds to average flux ratios as small as $F_{\rm  sec}/F_{\rm prim} \sim 10^{-3}$ and binary mass-ratios $M_{\rm  sec}/M_{\rm prim} \sim 0.2$, or a main-sequence M0 star orbiting an A0V primary.

The lowest detectable mass ratio is even more striking for young stars. At 1 Myr, both the A0 star and its companion are still contracting onto the main sequence \citep{Bressan2012}. The flux ratio limit corresponds to a $\sim$ M1 companion, similar to the main sequence case. However the mass ratio in this young system is $M_{\rm  sec}/M_{\rm prim} \sim 0.1$, half that of main-sequence components with similar spectral types. The direct spectral detection method is therefore well suited for finding close, low-mass companions to massive young stars.

There is also an upper detection limit near $6500\ K$ set by rotation. Our method of removing the primary star spectrum can also remove the companion spectrum if it has a similar rotational velocity, which hot companions are likely to have. Subtracting a model atmosphere for the primary star would remove the upper limit, but would reduce the detection rate for cool companions that are most difficult to detect with any other means.

Finally, we applied the direct spectral detection method to a set of known binary systems with close, late-type companions. We detected the companion spectrum in 9 of 34 known binary systems, 3 of which we characterized for the first time. Most of the companions we failed to detect are likely very cool, falling below the sensitivity limit of our data.

The direct spectral detection method is able to detect close binary companions with comparable or better sensitivity than imaging techniques, and does not require large telescopes with extremely competitive time allocation requests. This method is an excellent way to identify and perform initial characterization on new binary systems using smaller telescopes, but care must be taken to calibrate the parameter estimation. 

\section*{Acknowledgements}
This research has made use of the SIMBAD database, operated at CDS, Strasbourg, France, and of Astropy, a community-developed core Python package for Astronomy (Astropy Collaboration, 2013).
It was supported by a start-up grant to Adam Kraus as well as a University of Texas Continuing Fellowship to Kevin Gullikson. J.-E. Lee was supported by the Basic Science Research Program through the National Research Foundation of Korea (NRF) (grant No. NRF-2015R1A2A2A01004769) and the Korea
Astronomy and Space Science Institute under the R\&D program (Project No. 2015-1-320-18) supervised by the Ministry of Science, ICT and Future Planning.

This work used the Immersion Grating Infrared Spectrograph (IGRINS) that was developed under a collaboration between the University of Texas at Austin and the Korea Astronomy and Space Science Institute (KASI) with the financial support of the US National Science Foundation under grant AST-1229522, of the University of Texas at Austin, and of the Korean GMT Project of KASI.

The Hobby-Eberly Telescope (HET) is a joint project of the University of Texas at Austin, the Pennsylvania State University, Stanford University, Ludwig-Maximilians-Universit\"at M\"unchen, and Georg-August-Universit\"at G\"ottingen. The HET is named in honor of its principal benefactors, William P. Hobby and Robert E. Eberly.

Based on observations at Cerro Tololo Inter-American Observatory, National Optical Astronomy Observatory (NOAO Prop. IDs: 13A-0139, 13B-0112, 2014A-0260, 14A-0260, 15A-0245; PI: Kevin Gullikson), which is operated by the Association of Universities for Research in Astronomy (AURA) under a cooperative agreement with the National Science Foundation. 

We would like to thank Bill Cochran and Mike Endl for observing some of the spectra used in this project. . Finally, we would like to thank the anonymous referee for numerous comments that improved the paper.


\newpage
\clearpage

\LongTables
\begin{deluxetable*}{lllcccrcc}
\tabletypesize{\footnotesize}
\tablewidth{0pt}
\tablecaption{ Early type Calibration Stars \label{tab:earlycal}}
\tablehead{
\colhead{} & \colhead{} & \colhead{} & \colhead{}& \colhead{} & \colhead{} & \colhead{}  & \colhead{}  & \colhead{Exp. Time} \\
\colhead{Star} & \colhead{RA} & \colhead{DEC} & \colhead{SpT}& \colhead{V} & \colhead{K} & \colhead{Instrument}  & \colhead{Date}  & \colhead{(min)} }

\startdata

   HIP 1191 &  00:14:54.5 &  -09:34:10.4 &          B8.5V &     5.76 &     5.94 &     CHIRON &  2013-09-17 &          180.00 \\
    HIP 2381 &  00:30:22.6 &  -23:47:15.6 &            A3V &     5.19 &     4.83 &     CHIRON &  2014-08-05 &           58.18 \\
   HIP 10320 &  02:12:54.4 &  -30:43:25.7 &            B9V &     5.26 &     5.21 &     CHIRON &  2013-08-28 &          119.00 \\
   HIP 13717 &  02:56:37.4 &  -03:42:44.3 &            A3V &     5.16 &     4.86 &     CHIRON &  2014-11-09 &           74.99 \\
   HIP 14293 &  03:04:16.5 &  -07:36:03.0 &            A5V &     5.30 &     4.74 &     CHIRON &  2014-09-19 &           53.33 \\
   HIP 16285 &  03:29:55.1 &  -42:38:03.3 &            A5V &     5.77 &     5.18 &     CHIRON &  2014-10-03 &           76.29 \\
   HIP 17457 &  03:44:30.5 &  -01:09:47.1 &           B7IV &     5.25 &     5.43 &     CHIRON &  2013-08-27 &          107.13 \\
   HIP 18788 &  04:01:32.0 &  -01:32:58.7 &            B5V &     5.28 &     5.66 &     CHIRON &  2013-08-31 &          121.22 \\
   HIP 20264 &  04:20:39.0 &  -20:38:22.6 &            A0V &     5.38 &     5.33 &     CHIRON &  2014-03-02 &          100.33 \\
   HIP 20507 &  04:23:40.8 &  -03:44:43.6 &            A2V &     5.17 &     4.93 &     CHIRON &  2014-03-02 &           11.92 \\
   HIP 20507 &  04:23:40.8 &  -03:44:43.6 &            A2V &     5.17 &     4.93 &     CHIRON &  2014-03-03 &           52.30 \\
   HIP 22913 &  04:55:50.1 &  +15:02:25.0 &            B9V &     5.78 &     5.97 &     CHIRON &  2013-10-20 &          200.00 \\
   HIP 23362 &  05:01:25.5 &  -20:03:06.9 &            B9V &     4.89 &     4.97 &     CHIRON &  2013-09-13 &           84.58 \\
   HIP 25280 &  05:24:28.4 &  -16:58:32.8 &            A0V &     5.64 &     5.65 &     CHIRON &  2014-10-20 &           67.52 \\
   HIP 25608 &  05:28:15.3 &  -37:13:50.7 &            A1V &     5.56 &     5.50 &     CHIRON &  2014-03-02 &          103.00 \\
   HIP 27321 &  05:47:17.0 &  -51:03:59.4 &            A6V &     3.86 &     3.48 &     CHIRON &  2014-02-08 &           23.23 \\
   HIP 28910 &  06:06:09.3 &  -14:56:06.9 &            A0V &     4.67 &     4.52 &     CHIRON &  2014-02-05 &           33.31 \\
   HIP 29735 &  06:15:44.8 &  -13:43:06.2 &            B9V &     5.00 &     5.10 &     CHIRON &  2013-09-24 &           93.68 \\
   HIP 30069 &  06:19:40.9 &  -34:23:47.7 &            B9V &     5.75 &     5.93 &     CHIRON &  2013-10-08 &          180.00 \\
   HIP 30788 &  06:28:10.2 &  -32:34:48.2 &            B4V &     4.48 &     4.91 &     CHIRON &  2013-10-09 &           56.93 \\
   HIP 31362 &  06:34:35.3 &  -32:42:58.5 &            B8V &     5.61 &     5.73 &     CHIRON &  2013-11-02 &          160.00 \\
   HIP 32474 &  06:46:39.0 &  -10:06:26.4 &          B9.5V &     5.65 &     5.66 &     CHIRON &  2013-10-27 &          160.00 \\
   HIP 33575 &  06:58:35.8 &  -25:24:50.9 &            B2V &     5.58 &     6.05 &     CHIRON &  2013-11-03 &          140.00 \\
   HIP 35180 &  07:16:14.5 &  -15:35:08.4 &            A1V &     5.45 &     5.27 &     CHIRON &  2014-02-08 &           90.55 \\
     HR 2948 &  07:38:49.3 &  -26:48:06.4 &            B6V &     4.50 &     4.96 &     CHIRON &  2013-10-19 &           67.20 \\
   HIP 37450 &  07:41:15.8 &  -38:32:00.7 &            B5V &     5.41 &     5.78 &     CHIRON &  2013-11-04 &          136.62 \\
   HIP 40429 &  08:15:15.9 &  -62:54:56.3 &            A2V &     5.16 &  \nodata &     CHIRON &  2014-02-03 &           82.83 \\
   HIP 40706 &  08:18:33.3 &  -36:39:33.4 &            A8V &     4.40 &     4.00 &     CHIRON &  2013-02-04 &           32.32 \\
   HIP 42334 &  08:37:52.1 &  -26:15:18.0 &            A0V &     5.27 &     5.32 &     CHIRON &  2014-02-24 &           48.60 \\
   HIP 45344 &  09:14:24.4 &  -43:13:38.9 &            B4V &     5.25 &     5.59 &     CHIRON &  2013-11-16 &          116.78 \\
     HR 4259 &  10:55:36.8 &  +24:44:59.0 &            A1V &     4.50 &  \nodata &     CHIRON &  2013-02-12 &           35.47 \\
   HIP 56633 &  11:36:40.9 &  -09:48:08.0 &         B9.5Vn &     4.68 &     4.78 &     CHIRON &  2013-02-12 &           41.77 \\
   HIP 57328 &  11:45:17.0 &  +08:15:29.2 &            A4V &     4.84 &     4.41 &     CHIRON &  2013-02-15 &           48.88 \\
   HIP 57328 &  11:45:17.0 &  +08:15:29.2 &            A4V &     4.84 &     4.41 &     CHIRON &  2013-03-19 &           48.88 \\
   HIP 61622 &  12:37:42.1 &  -48:32:28.6 &         A1IVnn &     3.86 &     3.70 &     CHIRON &  2013-03-27 &           19.48 \\
   HIP 66249 &  13:34:41.7 &  -00:35:45.3 &          A2Van &     3.38 &     3.07 &     CHIRON &  2013-03-27 &           12.83 \\
   HIP 66821 &  13:41:44.7 &  -54:33:33.9 &  B8.5Vn        &     5.01 &  \nodata &     CHIRON &  2014-03-02 &           71.17 \\
   HIP 68520 &  14:01:38.7 &  +01:32:40.3 &            A3V &     4.24 &     4.09 &     CHIRON &  2013-04-21 &           27.88 \\
   HIP 70327 &  14:23:22.6 &  +08:26:47.8 &            A0V &     5.12 &     5.07 &     CHIRON &  2014-03-03 &           42.15 \\
   HIP 72104 &  14:44:59.2 &  -35:11:30.5 &            A0V &     4.92 &     4.78 &     CHIRON &  2014-03-04 &           58.09 \\
   HIP 73049 &  14:55:44.7 &  -33:51:20.8 &            A0V &     5.32 &     5.13 &     CHIRON &  2014-02-27 &           69.75 \\
   HIP 75304 &  15:23:09.3 &  -36:51:30.5 &            B4V &     4.54 &     4.94 &     CHIRON &  2013-05-15 &           36.05 \\
   HIP 77233 &  15:46:11.2 &  +15:25:18.5 &            A3V &     3.67 &     3.42 &     CHIRON &  2013-05-14 &           16.33 \\
   HIP 77635 &  15:50:58.7 &  -25:45:04.6 &         B1.5Vn &     4.64 &     4.78 &     CHIRON &  2014-03-09 &           30.80 \\
   HIP 78105 &  15:56:53.4 &  -33:57:58.0 &            A3V &     5.08 &     4.85 &     CHIRON &  2014-07-31 &           21.30 \\
   HIP 78105 &  15:56:53.4 &  -33:57:58.0 &            A3V &     5.08 &     4.85 &     CHIRON &  2014-08-01 &           48.91 \\
   HIP 78106 &  15:56:54.1 &  -33:57:51.3 &            B9V &     5.55 &     5.42 &     CHIRON &  2014-03-20 &           70.48 \\
   HIP 78554 &  16:02:17.6 &  +22:48:16.0 &            A3V &     4.82 &     4.62 &     CHIRON &  2013-05-15 &           47.60 \\
   HIP 79007 &  16:07:37.5 &  +09:53:30.2 &            A7V &     5.64 &     5.09 &     CHIRON &  2014-08-04 &           58.79 \\
   HIP 79007 &  16:07:37.5 &  +09:53:30.2 &            A7V &     5.64 &     5.09 &     CHIRON &  2014-08-05 &           21.23 \\
   HIP 79387 &  16:12:07.3 &  -08:32:51.2 &            A4V &     5.43 &     5.05 &     CHIRON &  2014-03-30 &           70.72 \\
   HIP 79653 &  16:15:15.3 &  -47:22:19.2 &            B8V &     5.12 &     5.42 &     CHIRON &  2014-03-24 &           47.12 \\
   HIP 80815 &  16:30:12.4 &  -25:06:54.8 &            B3V &     4.79 &     5.10 &     CHIRON &  2013-03-27 &           45.85 \\
   HIP 85537 &  17:28:49.6 &  +00:19:50.2 &            A7V &     5.42 &     4.80 &     CHIRON &  2014-05-15 &           60.83 \\
   HIP 85922 &  17:33:29.8 &  -05:44:41.2 &            A5V &     5.62 &     5.14 &     CHIRON &  2014-08-17 &           95.90 \\
   HIP 86019 &  17:34:46.3 &  -11:14:31.1 &           B8Vn &     5.54 &     5.36 &     CHIRON &  2014-03-31 &           69.76 \\
   HIP 87108 &  17:47:53.5 &  +02:42:26.2 &        A1Vnk &     3.75 &     3.65 &     CHIRON &  2013-06-02 &           17.73 \\
   HIP 90887 &  18:32:21.3 &  -39:42:14.4 &           A3Vn &     5.16 &     4.93 &     CHIRON &  2014-04-01 &           73.61 \\
   HIP 91875 &  18:43:46.9 &  -38:19:24.3 &           A2Vn &     5.12 &     4.86 &     CHIRON &  2014-03-29 &           46.61 \\
   HIP 92946 &  18:56:13.1 &  +04:12:12.9 &            A5V &     4.62 &     4.09 &     CHIRON &  2013-07-02 &           39.55 \\
   HIP 93805 &  19:06:14.9 &  -04:52:57.2 &           B9Vn &     3.43 &     3.65 &     CHIRON &  2014-04-28 &           10.92 \\
  HIP 101589 &  20:35:18.5 &  +14:40:27.1 &            A3V &     4.66 &     4.36 &     CHIRON &  2013-06-05 &           41.07 \\
  HIP 104139 &  21:05:56.8 &  -17:13:58.3 &            A1V &     4.07 &     4.10 &     CHIRON &  2013-06-05 &           23.80 \\
  HIP 105140 &  21:17:56.2 &  -32:10:21.1 &            A1V &     4.72 &     4.49 &     CHIRON &  2013-07-12 &           43.40 \\
  HIP 107517 &  21:46:32.0 &  -11:21:57.4 &            A1V &     5.57 &     5.57 &     CHIRON &  2014-08-04 &          118.70 \\
  HIP 107608 &  21:47:44.1 &  -30:53:53.9 &            A2V &     5.02 &     4.85 &     CHIRON &  2014-05-11 &           52.76 \\
  HIP 108294 &  21:56:22.7 &  -37:15:13.1 &           A2Vn &     5.46 &     5.17 &     CHIRON &  2014-05-13 &           57.20 \\
  HIP 110935 &  22:28:37.6 &  -67:29:20.6 &            A4V &     5.57 &     5.05 &     CHIRON &  2014-08-27 &           74.48 \\
  HIP 117089 &  23:44:12.0 &  -18:16:36.9 &            B9V &     5.24 &     5.38 &     CHIRON &  2013-08-09 &          102.52 \\
    HIP 5361 &  01:08:33.4 &  +58:15:48.4 &            B8V &     5.77 &     5.75 &        HRS &  2013-08-19 &           50.00 \\
    HIP 8016 &  01:42:55.8 &  +70:37:21.0 &            B9V &     5.18 &     5.22 &        HRS &  2013-08-18 &           16.40 \\
   HIP 14043 &  03:00:52.2 &  +52:21:06.2 &            B7V &     5.25 &     5.43 &        HRS &  2013-08-19 &           20.00 \\
   HIP 14143 &  03:02:22.5 &  +04:21:10.3 &            B7V &     5.61 &     5.90 &        HRS &  2013-08-14 &           23.10 \\
   HIP 15404 &  03:18:37.7 &  +50:13:19.8 &            B3V &     5.16 &     5.33 &        HRS &  2013-08-13 &           10.25 \\
   HIP 18396 &  03:55:58.1 &  +47:52:17.1 &            B6V &     5.38 &     5.58 &        HRS &  2013-08-12 &           12.95 \\
   HIP 20430 &  04:22:34.9 &  +25:37:45.5 &          B9Vnn &     5.38 &     5.45 &        HRS &  2013-08-16 &           18.00 \\
   HIP 20579 &  04:24:29.1 &  +34:07:50.7 &            B8V &     5.72 &     5.81 &        HRS &  2013-08-13 &           24.50 \\
   HIP 66798 &  13:41:29.8 &  +64:49:20.6 &            A2V &     5.85 &     5.65 &        HRS &  2013-03-26 &           18.10 \\
   HIP 67194 &  13:46:13.5 &  +41:05:19.4 &            A5V &     5.89 &     5.34 &        HRS &  2013-04-07 &           18.40 \\
   HIP 67782 &  13:53:10.2 &  +28:38:53.2 &            A7V &     5.91 &     5.47 &        HRS &  2013-04-12 &           17.95 \\
   HIP 70384 &  14:24:00.8 &  +08:14:38.2 &            A3V &     5.93 &     5.72 &        HRS &  2013-04-21 &           21.00 \\
   HIP 72154 &  14:45:30.2 &  +00:43:02.1 &          B9.5V &     5.67 &     5.60 &        HRS &  2013-04-21 &           15.00 \\
   HIP 80991 &  16:32:25.6 &  +60:49:23.9 &            A2V &     5.91 &     5.78 &        HRS &  2013-04-07 &           20.20 \\
   HIP 82350 &  16:49:34.6 &  +13:15:40.1 &            A1V &     5.91 &     5.86 &        HRS &  2013-04-09 &           20.20 \\
   HIP 83635 &  17:05:32.2 &  -00:53:31.4 &            B1V &     5.61 &     5.29 &        HRS &  2013-04-25 &           15.80 \\
   HIP 85379 &  17:26:44.2 &  +48:15:36.2 &            A4V &     5.83 &     5.38 &        HRS &  2013-04-16 &           16.50 \\
   HIP 86782 &  17:43:59.1 &  +53:48:06.1 &            A2V &     5.76 &     5.59 &        HRS &  2013-04-22 &           15.50 \\
   HIP 88817 &  18:07:49.5 &  +26:05:50.4 &            A3V &     5.90 &     5.51 &        HRS &  2013-04-23 &           20.00 \\
   HIP 90052 &  18:22:35.3 &  +12:01:46.8 &            A2V &     5.98 &     5.77 &        HRS &  2013-04-23 &           25.00 \\
   HIP 92312 &  18:48:53.3 &  +19:19:43.3 &            A1V &     5.89 &     5.82 &        HRS &  2013-04-26 &           19.50 \\
   HIP 93393 &  19:01:17.3 &  +26:17:29.0 &            B5V &     5.68 &     5.84 &        HRS &  2013-04-22 &           16.00 \\
   HIP 96840 &  19:41:05.5 &  +13:48:56.4 &            B5V &     5.99 &     6.21 &        HRS &  2013-04-26 &           27.60 \\
  HIP 100069 &  20:18:06.9 &  +40:43:55.5 &            O9V &     5.84 &     5.72 &        HRS &  2013-04-27 &           22.55 \\
  HIP 105282 &  21:19:28.7 &  +49:30:37.0 &            B6V &     5.74 &     6.08 &        HRS &  2013-08-18 &           36.77 \\
  HIP 105942 &  21:27:21.3 &  +37:07:00.4 &            B3V &     5.29 &     5.64 &        HRS &  2013-08-19 &           24.00 \\
  HIP 105972 &  21:27:46.1 &  +66:48:32.7 &            B7V &     5.41 &     5.60 &        HRS &  2013-08-03 &           13.35 \\
    HIP 5132 &  01:05:41.7 &  +21:27:55.5 &           A0Vn &     5.53 &     5.61 &     IGRINS &  2014-07-09 &            6.67 \\
    HIP 5518 &  01:10:39.3 &  +68:46:43.0 &          A0Vnn &     5.32 &     5.31 &     IGRINS &  2014-10-15 &            3.73 \\
    HIP 5626 &  01:12:16.8 &  +79:40:26.2 &            A3V &     5.60 &     5.49 &     IGRINS &  2014-10-15 &            3.73 \\
    HIP 9564 &  02:02:52.4 &  +64:54:05.2 &           A1Vn &     6.00 &     5.92 &     IGRINS &  2014-10-15 &            3.73 \\
   HIP 12803 &  02:44:32.9 &  +15:18:42.7 &           B9Vn &     5.78 &     5.79 &     IGRINS &  2014-10-17 &            3.73 \\
   HIP 13879 &  02:58:45.6 &  +39:39:45.8 &           A2Vn &     4.70 &     4.42 &     IGRINS &  2014-10-15 &            3.73 \\
   HIP 14862 &  03:11:56.2 &  +74:23:37.1 &          A2Vnn &     4.84 &     4.71 &     IGRINS &  2014-10-15 &            3.73 \\
   HIP 15110 &  03:14:54.0 &  +21:02:40.0 &            A1V &     4.88 &     4.82 &     IGRINS &  2014-10-16 &            4.20 \\
   HIP 16599 &  03:33:39.0 &  +54:58:29.4 &            A3V &     5.98 &     5.68 &     IGRINS &  2014-10-15 &            3.73 \\
   HIP 17527 &  03:45:09.7 &  +24:50:21.3 &            B8V &     5.64 &     5.81 &     IGRINS &  2014-10-17 &            3.73 \\
   HIP 20789 &  04:27:17.4 &  +22:59:46.8 &            B7V &     5.51 &     5.74 &     IGRINS &  2014-10-16 &            3.73 \\
   HIP 21683 &  04:39:16.5 &  +15:55:04.7 &           A5Vn &     4.68 &     4.23 &     IGRINS &  2014-10-18 &            3.83 \\
   HIP 22028 &  04:44:07.9 &  -18:39:59.7 &            A1V &     5.53 &     5.44 &     IGRINS &  2014-10-17 &            4.00 \\
   HIP 23362 &  05:01:25.5 &  -20:03:06.9 &            B9V &     4.89 &     4.97 &     IGRINS &  2014-10-16 &            4.00 \\
 ADS 3962 AB &  05:22:50.3 &     +03 32 52 &           B1Vn &     4.99 &  \nodata &     IGRINS &  2014-10-16 &            4.67 \\
   HIP 25143 &  05:22:50.3 &  +41:01:45.3 &            A3V &     5.55 &     5.11 &     IGRINS &  2014-10-16 &            3.73 \\
   HIP 25280 &  05:24:28.4 &  -16:58:32.8 &            A0V &     5.64 &     5.65 &     IGRINS &  2014-10-17 &            4.00 \\
   HIP 25790 &  05:30:26.1 &  +15:21:37.6 &           A3Vn &     5.94 &     5.55 &     IGRINS &  2014-10-16 &            3.73 \\
   HIP 26093 &  05:33:54.2 &  +14:18:20.0 &            B3V &     5.59 &     5.96 &     IGRINS &  2014-10-16 &            4.67 \\
   HIP 27713 &  05:52:07.7 &  -09:02:30.8 &           A2Vn &     5.96 &     5.65 &     IGRINS &  2014-10-16 &            4.00 \\
   HIP 29151 &  06:08:57.9 &  +02:29:58.8 &           A3Vn &     5.73 &     5.35 &     IGRINS &  2014-10-16 &            4.40 \\
   HIP 29735 &  06:15:44.8 &  -13:43:06.2 &            B9V &     5.00 &     5.10 &     IGRINS &  2014-10-16 &            4.00 \\
   HIP 30666 &  06:26:39.5 &  -01:30:26.4 &           A3Vn &     5.87 &     5.64 &     IGRINS &  2014-10-16 &            4.67 \\
   HIP 31278 &  06:33:37.9 &  -01:13:12.5 &           B5Vn &     5.08 &     5.46 &     IGRINS &  2014-10-16 &            4.00 \\
   HIP 36812 &  07:34:15.8 &  +03:22:18.1 &          A0Vnn &     5.83 &     5.74 &     IGRINS &  2014-10-17 &            4.00 \\
   HIP 40881 &  08:20:32.1 &  +24:01:20.3 &          B9.5V &     5.93 &     5.91 &     IGRINS &  2014-10-17 &            4.00 \\
   HIP 85290 &  17:25:41.3 &  +60:02:54.2 &           A1Vn &     5.64 &     5.50 &     IGRINS &  2014-10-16 &            3.73 \\
   HIP 85385 &  17:26:49.1 &  +20:04:51.5 &            B5V &     5.51 &     5.84 &     IGRINS &  2014-07-10 &            8.00 \\
   HIP 93713 &  19:04:55.1 &  +53:23:47.9 &           A0Vn &     5.38 &     5.41 &     IGRINS &  2014-07-10 &            8.00 \\
   HIP 94620 &  19:15:17.3 &  +21:13:55.6 &            A4V &     5.65 &     5.30 &     IGRINS &  2014-07-10 &            10.00 \\
   HIP 97376 &  19:47:27.7 &  +38:24:27.4 &           B8Vn &     5.83 &     6.01 &     IGRINS &  2014-07-10 &            8.00 \\
   HIP 99742 &  20:14:16.6 &  +15:11:51.3 &            A2V &     4.95 &     4.77 &     IGRINS &  2014-10-15 &            8.00 \\
  HIP 101123 &  20:29:53.9 &  -18:34:59.4 &            A1V &     5.91 &     5.72 &     IGRINS &  2014-10-15 &            4.00 \\
  HIP 101909 &  20:39:04.9 &  +15:50:17.5 &            B3V &     5.98 &  \nodata &     IGRINS &  2014-10-15 &            6.00 \\
  HIP 102487 &  20:46:09.9 &  -21:30:50.5 &            A1V &     5.91 &     5.77 &     IGRINS &  2014-07-09 &            8.00 \\
  HIP 104365 &  21:08:33.6 &  -21:11:37.2 &            A0V &     5.28 &     5.30 &     IGRINS &  2014-07-09 &            8.00 \\
  HIP 105891 &  21:26:44.9 &  +52:53:54.7 &          B7III &     5.99 &     6.34 &     IGRINS &  2014-10-16 &            3.73 \\
  HIP 108339 &  21:56:56.3 &  +12:04:35.3 &          A2Vnn &     5.54 &     5.36 &     IGRINS &  2014-10-15 &            3.73 \\
  HIP 109831 &  22:14:44.3 &  +42:57:14.0 &          A2Vnn &     5.72 &     5.66 &     IGRINS &  2014-10-15 &            3.73 \\
  HIP 111056 &  22:29:52.9 &  +78:49:27.4 &            A3V &     5.46 &     5.23 &     IGRINS &  2014-10-15 &            4.67 \\
    HIP 1366 &  00:17:05.4 &  +38:40:53.8 &            A2V &     4.61 &     4.42 &       TS23 &  2013-10-20 &           32.17 \\
    HIP 4436 &  00:56:45.2 &  +38:29:57.6 &            A5V &     3.87 &     3.49 &       TS23 &  2013-10-20 &           18.14 \\
    HIP 9312 &  01:59:38.0 &  +64:37:17.7 &           A0Vn &     5.28 &     5.22 &       TS23 &  2013-10-21 &           59.25 \\
   HIP 13327 &  02:51:29.5 &  +15:04:55.4 &            B7V &     5.51 &     5.78 &       TS23 &  2014-01-13 &          120.70 \\
   HIP 15444 &  03:19:07.6 &  +50:05:41.8 &            B5V &     5.04 &     5.20 &       TS23 &  2013-10-17 &           49.84 \\
   HIP 16340 &  03:30:36.9 &  +48:06:12.9 &            B8V &     5.82 &     5.90 &       TS23 &  2014-01-21 &           71.58 \\
   HIP 18141 &  03:52:41.6 &  -05:21:40.5 &            B8V &     5.48 &     5.71 &       TS23 &  2014-01-21 &           58.26 \\
   HIP 21819 &  04:41:19.7 &  +28:36:53.9 &            A2V &     5.73 &     5.70 &       TS23 &  2014-01-22 &           74.02 \\
   HIP 21928 &  04:42:54.3 &  +43:21:54.5 &           A1Vn &     5.30 &     5.20 &       TS23 &  2014-01-20 &           73.64 \\
   HIP 25555 &  05:27:45.6 &  +15:52:26.5 &         B9.5Vn &     5.51 &     5.33 &       TS23 &  2014-01-13 &           95.73 \\
   HIP 29997 &  06:18:50.7 &  +69:19:11.2 &           A0Vn &     4.76 &     4.67 &       TS23 &  2014-01-22 &           35.16 \\
   HIP 31434 &  06:35:12.0 &  +28:01:20.3 &          A0Vnn &     5.27 &     5.15 &       TS23 &  2014-01-19 &           58.71 \\
   HIP 34769 &  07:11:51.8 &  -00:29:33.9 &            A2V &     4.15 &     3.90 &       TS23 &  2014-01-20 &           27.36 \\
   HIP 35341 &  07:18:02.2 &  +40:53:00.2 &           A5Vn &     5.87 &     5.33 &       TS23 &  2014-01-23 &           83.54 \\
   HIP 36393 &  07:29:20.4 &  +28:07:05.7 &            A4V &     5.07 &     4.74 &       TS23 &  2014-01-19 &           51.08 \\
   HIP 38538 &  07:53:29.8 &  +26:45:56.8 &            A3V &     4.98 &     4.66 &       TS23 &  2014-01-12 &           56.54 \\
   HIP 39236 &  08:01:30.2 &  +16:27:19.1 &         B9.5Vn &     5.99 &     5.94 &       TS23 &  2014-01-22 &          128.84 \\
   HIP 41307 &  08:25:39.6 &  -03:54:23.1 &            A0V &     3.90 &     3.93 &       TS23 &  2014-01-10 &           43.84 \\
   HIP 42313 &  08:37:39.3 &  +05:42:13.6 &          A1Vnn &     4.14 &     4.03 &       TS23 &  2014-01-24 &           58.27 \\
   HIP 43142 &  08:47:14.9 &  -01:53:49.3 &            A3V &     5.28 &     5.04 &       TS23 &  2014-01-13 &           83.95 \\
   HIP 44127 &  08:59:12.4 &  +48:02:30.5 &         A7V(n) &     3.14 &     2.66 &       TS23 &  2014-01-20 &           18.24 \\
   HIP 47006 &  09:34:49.4 &  +52:03:05.3 &           A0Vn &     4.48 &     4.34 &       TS23 &  2014-01-19 &           27.62 \\
   HIP 50303 &  10:16:14.4 &  +29:18:37.8 &           A0Vn &     5.49 &     5.39 &       TS23 &  2014-01-20 &          116.86 \\
   HIP 50860 &  10:23:06.3 &  +33:54:29.3 &            A6V &     5.90 &     5.51 &       TS23 &  2014-01-21 &          138.85 \\
   HIP 51685 &  10:33:30.9 &  +34:59:19.2 &           A2Vn &     5.58 &     5.35 &       TS23 &  2014-01-20 &           92.78 \\
   HIP 52422 &  10:43:01.8 &  +26:19:32.0 &           A4Vn &     5.52 &     5.05 &       TS23 &  2014-01-19 &           52.89 \\
   HIP 52457 &  10:43:24.9 &  +23:11:18.2 &           A3Vn &     5.07 &     4.92 &       TS23 &  2014-01-19 &           44.36 \\
   HIP 52638 &  10:45:51.8 &  +30:40:56.3 &           A1Vn &     5.35 &     5.40 &       TS23 &  2014-01-12 &           94.63 \\
   HIP 52911 &  10:49:15.4 &  +10:32:42.7 &            A2V &     5.31 &     5.07 &       TS23 &  2014-01-13 &           99.47 \\
   HIP 54849 &  11:13:45.5 &  -00:04:10.2 &            A0V &     5.40 &     5.33 &       TS23 &  2014-01-13 &          150.85 \\
   HIP 56034 &  11:29:04.1 &  +39:20:13.1 &            A2V &     5.35 &     5.31 &       TS23 &  2014-01-19 &           39.60 \\
   HIP 59819 &  12:16:00.1 &  +14:53:56.6 &            A3V &     5.09 &     4.89 &       TS23 &  2014-01-12 &           67.54 \\
   HIP 60595 &  12:25:11.7 &  -11:36:38.1 &            A1V &     5.95 &     5.83 &       TS23 &  2014-01-19 &          114.15 \\
   HIP 60957 &  12:29:43.2 &  +20:53:45.9 &            A3V &     5.68 &     5.43 &       TS23 &  2014-01-21 &           92.45 \\
   HIP 65728 &  13:28:27.0 &  +59:56:44.8 &           A1Vn &     5.40 &     5.43 &       TS23 &  2014-01-20 &          106.14 \\
   HIP 75178 &  15:21:48.5 &  +32:56:01.3 &           B9Vn &     5.38 &     5.49 &       TS23 &  2014-01-21 &           84.18 \\
   HIP 93747 &  19:05:24.6 &  +13:51:48.5 &       A0Vnn &     2.99 &     2.88 &       TS23 &  2013-10-22 &           10.90 \\
   HIP 95853 &  19:29:42.3 &  +51:43:47.2 &            A5V &     3.77 &     3.60 &       TS23 &  2013-10-20 &           18.37 \\
   HIP 96288 &  19:34:41.2 &  +42:24:45.0 &            A2V &     5.35 &     5.05 &       TS23 &  2013-10-20 &           67.76 \\
   HIP 99080 &  20:06:53.4 &  +23:36:51.9 &            B3V &     5.06 &     5.57 &       TS23 &  2013-10-18 &           55.24 \\
  HIP 101716 &  20:37:04.6 &  +26:27:43.0 &            B8V &     5.59 &     5.71 &       TS23 &  2013-10-17 &           49.96 \\
  HIP 105966 &  21:27:40.0 &  +27:36:30.9 &            A1V &     5.39 &     5.29 &       TS23 &  2013-10-20 &           72.23 \\
  HIP 111169 &  22:31:17.5 &  +50:16:56.9 &            A1V &     3.77 &     3.75 &       TS23 &  2013-10-20 &           17.03 \\
  HIP 111841 &  22:39:15.6 &  +39:03:00.9 &            O9V &     4.88 &     5.50 &       TS23 &  2013-10-18 &           35.70 \\
  HIP 113788 &  23:02:36.3 &  +42:45:28.0 &           A3Vn &     5.10 &     4.69 &       TS23 &  2013-10-21 &           47.51 \\
  HIP 114520 &  23:11:44.1 &  +08:43:12.3 &           A5Vn &     5.16 &     4.74 &       TS23 &  2013-10-22 &           72.27 \\
  HIP 117371 &  23:47:54.7 &  +67:48:24.5 &           A1Vn &     5.05 &     4.97 &       TS23 &  2013-10-21 &           44.52 \\

\enddata
\tablecomments{The spectral types are from the Simbad database \citep{Simbad}.}
\end{deluxetable*}

\LongTables
\begin{deluxetable*}{lcccccrccc}
\tabletypesize{\footnotesize}
\tablewidth{0pt}
\tablecaption{ Late type Calibration Stars \label{tab:latecal}}
\tablehead{
\colhead{} & \colhead{} & \colhead{} & \colhead{} & \colhead{} & \colhead{} & \colhead{}  & \colhead{} & \colhead{Exp. Time} \\
\colhead{Star} & \colhead{RA} & \colhead{DEC} & \colhead{V} & \colhead{K} & \colhead{$\rm T_{eff}$ (K)} & \colhead{Instrument}  & \colhead{Date} & \colhead{(min)} }

\startdata

   HD 33793 &   05:11:40.5 &   -45:01:06.2 &   8.85 &  5.05 &    $3570 \pm 160^{1}$ &      CHIRON &      2015-01-13 &             60.00  \\
   HD 36379 &   05:30:59.9 &   -10:04:51.9 &   6.91 &  5.56 &     $6030 \pm 14^{2}$ &      CHIRON &      2015-01-14 &              9.58  \\
   HD 38858 &   05:48:34.9 &   -04:05:40.7 &   5.97 &  4.41 &     $5646 \pm 45^{3}$ &      CHIRON &      2015-01-14 &              5.31  \\
   HD 42581 &   06:10:34.6 &   -21:51:52.7 &   8.12 &  4.17 &    $3814 \pm 113^{4}$ &      CHIRON &      2015-01-14 &             30.62  \\
   HD 45184 &   06:24:43.8 &   -28:46:48.4 &   6.39 &  4.87 &     $5869 \pm 14^{2}$ &      CHIRON &      2015-01-14 &              5.30  \\
   HD 50806 &   06:53:33.9 &   -28:32:23.2 &   6.04 &  4.33 &     $5633 \pm 15^{2}$ &      CHIRON &      2015-01-14 &              3.99  \\
   HD 61421 &   07:39:18.1 &   +05:13:29.9 &   0.37 & -0.65 &     $6582 \pm 16^{3}$ &      CHIRON &      2015-01-16 &              0.05  \\
   HD 69830 &   08:18:23.9 &   -12:37:55.8 &   5.95 &  4.16 &     $5402 \pm 28^{2}$ &      CHIRON &      2015-01-14 &              5.03  \\
   HD 102634 &   11:49:01.2 &   -00:19:07.2 &   6.15 &  4.92 &     $6215 \pm 44^{5}$ &      CHIRON &      2015-01-17 &              5.18  \\
     GJ 465 &   12:24:52.5 &   -18:14:32.2 &  11.27 &  6.95 &    $3472 \pm 110^{6}$ &      CHIRON &      2015-01-17 &             65.00  \\
   HD 115617 &   13:18:24.3 &   -18:18:40.3 &   4.74 &  2.96 &     $5558 \pm 19^{2}$ &      CHIRON &      2015-01-17 &              1.32  \\
   HD 125072 &   14:19:04.8 &   -59:22:44.5 &   6.66 &  4.33 &     $4903 \pm 44^{5}$ &      CHIRON &      2015-02-11 &              9.14  \\
   HD 128621 &   14:39:35.0 &   -60:50:15.0 &   1.33 & -0.60 &      $5232 \pm 8^{3}$ &      CHIRON &      2015-02-06 &              0.03  \\
  HD 154363 &   17:05:03.3 &   -05:03:59.4 &   7.71 &  4.73 &     $4723 \pm 89^{2}$ &      CHIRON &      2015-03-12 &             26.27  \\
  HD 157881 &   17:25:45.2 &   +02:06:41.1 &   7.56 &  4.14 &     $4124 \pm 60^{7}$ &      CHIRON &      2015-03-13 &             25.86  \\
   HD 165222 &   18:05:07.5 &   -03:01:52.7 &   9.36 &  5.31 &     $3416 \pm 40^{7}$ &      CHIRON &      2015-02-11 &              3.83  \\
  HD 225239 &   00:04:53.7 &   +34:39:35.2 &   6.11 &  4.44 &     $5699 \pm 80^{8}$ &         HRS &      2002-09-18 &              8.00  \\
    HD 3651 &   00:39:21.8 &   +21:15:01.7 &   5.88 &  4.00 &     $5046 \pm 86^{3}$ &         HRS &      2005-07-30 &              3.00  \\
   HD 16895 &   02:44:11.9 &   +49:13:42.4 &   4.11 &  2.78 &     $6344 \pm 44^{5}$ &         HRS &      2006-12-02 &              0.11  \\
   HD 38529 &   05:46:34.9 &   +01:10:05.4 &   5.94 &  4.21 &     $5697 \pm 44^{5}$ &         HRS &      2004-12-02 &              0.55  \\
     GJ 270 &   07:19:31.2 &   +32:49:48.3 &  10.05 &  6.38 &     $3668 \pm 54^{9}$ &         HRS &      2002-12-11 &             20.00  \\
   HD 58855 &   07:29:55.9 &   +49:40:20.8 &   5.36 &  4.18 &     $6398 \pm 80^{8}$ &         HRS &      2006-03-12 &              0.50  \\
     GJ 281 &   07:39:23.0 &   +02:11:01.1 &   9.59 &  5.87 &   $3776 \pm 145^{10}$ &         HRS &      2003-01-19 &             20.00  \\
   HD 69056 &   08:15:33.2 &   +11:25:51.4 &   7.70 &  6.06 &     $5635 \pm 55^{8}$ &         HRS &      2003-12-02 &             13.00  \\
   HD 73732 &   08:52:35.8 &   +28:19:50.9 &   5.95 &  4.01 &     $5235 \pm 44^{5}$ &         HRS &      2003-10-15 &              3.33  \\
     GJ 328 &   08:55:07.5 &   +01:32:56.4 &   9.98 &  6.35 &   $3828 \pm 168^{10}$ &         HRS &      2003-01-14 &             20.00  \\
   HD 79969 &   09:17:53.4 &   +28:33:37.8 &   7.21 &  4.77 &     $4825 \pm 8^{11}$ &         HRS &      2003-12-02 &             10.00  \\
 HIP 53070 &   10:51:28.1 &   +20:16:38.9 &   8.22 &  6.83 &     $6110 \pm 76^{8}$ &         HRS &      2009-02-14 &             20.00  \\
 HIP 53169 &   10:52:36.4 &   -02:06:33.5 &   9.82 &  7.05 &    $4525 \pm 47^{12}$ &         HRS &      2009-01-09 &             15.00  \\
      GJ 411 &   11:03:20.1 &   +35:58:11.5 &   7.52 &  3.34 &    $3464 \pm 15^{13}$ &         HRS &      2001-12-27 &              5.00  \\
  HD 114783 &   13:12:43.7 &   -02:15:54.1 &   7.55 &  5.47 &     $5135 \pm 44^{5}$ &         HRS &      2005-01-08 &              7.08  \\
     GJ 525 &   13:45:05.0 &   +17:47:07.5 &   9.75 &  6.22 &   $3680 \pm 150^{14}$ &         HRS &      2008-04-21 &             15.00  \\
     GJ 535 &   13:59:19.4 &   +22:52:11.1 &   9.04 &  6.24 &     $4580 \pm 7^{11}$ &         HRS &      2002-04-29 &             12.16  \\
  HD 142267 &   15:53:12.0 &   +13:11:47.8 &   6.12 &  4.53 &     $5756 \pm 44^{5}$ &         HRS &      2002-08-11 &              2.08  \\
     GJ 687 &   17:36:25.8 &   +68:20:20.9 &   9.15 &  4.55 &    $3413 \pm 28^{13}$ &         HRS &      2002-04-30 &             12.50  \\
     GJ 699 &   17:57:48.4 &   +04:41:36.2 &   9.51 &  4.52 &    $3222 \pm 10^{13}$ &         HRS &      2002-05-25 &             15.00  \\
     GJ 699 &   17:57:48.4 &   +04:41:36.2 &   9.51 &  4.52 &    $3222 \pm 10^{13}$ &         HRS &      2002-05-25 &             35.00  \\
     GL 15A &   00:18:22.8 &   +44:01:22.6 &   8.13 &  4.02 &    $3567 \pm 11^{13}$ &      IGRINS &      2014-11-23 &              4.00  \\
     GL 15B &   00:18:25.4 &   +44:01:37.6 &  11.04 &  5.95 &     $3218 \pm 60^{7}$ &      IGRINS &      2014-11-23 &              8.00  \\
     HD 1835 &   00:22:51.7 &   -12:12:33.9 &   6.39 &  4.86 &     $5837 \pm 44^{5}$ &      IGRINS &      2014-12-07 &              8.00  \\
    HD 4614 &   00:49:06.2 &   +57:48:54.6 &   3.44 &  1.99 &      $5973 \pm 8^{3}$ &      IGRINS &      2014-12-06 &              0.67  \\
    HD 10476 &   01:42:29.7 &   +20:16:06.6 &   5.24 &  3.25 &     $5242 \pm 12^{3}$ &      IGRINS &      2014-11-18 &              2.67  \\
    GL 1094 &   07:02:42.9 &   -06:47:57.2 &   8.35 &  5.76 &     $4698 \pm 91^{2}$ &      IGRINS &      2014-11-24 &              6.00  \\
    HD 58946 &   07:29:06.7 &   +31:47:04.3 &   4.18 &  2.98 &     $6597 \pm 18^{3}$ &      IGRINS &      2015-01-20 &              0.83  \\
    HD 67767 &   08:10:27.1 &   +25:30:26.4 &   5.73 &  3.84 &     $5344 \pm 44^{5}$ &      IGRINS &      2015-01-20 &              1.50  \\
    HD 71148 &   08:27:36.7 &   +45:39:10.7 &   6.30 &  4.83 &     $5818 \pm 44^{5}$ &      IGRINS &      2015-01-20 &              4.67  \\
   HD 76151 &   08:54:17.9 &   -05:26:04.0 &   6.00 &  4.46 &     $5788 \pm 23^{2}$ &      IGRINS &      2014-11-23 &              8.00  \\
    HD 87141 &   10:04:36.3 &   +53:53:30.1 &   5.72 &  4.50 &     $6401 \pm 80^{8}$ &      IGRINS &      2015-01-23 &              4.67  \\
    HD 87822 &   10:08:15.8 &   +31:36:14.5 &   6.24 &  5.13 &     $6586 \pm 80^{8}$ &      IGRINS &      2015-01-23 &             26.67  \\
    HD 91752 &   10:36:21.4 &   +36:19:36.9 &   6.30 &  5.20 &     $6543 \pm 80^{8}$ &      IGRINS &      2015-01-20 &             24.00  \\
    HD 95128 &   10:59:27.9 &   +40:25:48.9 &   5.04 &  3.75 &     $5882 \pm 44^{5}$ &      IGRINS &      2015-01-23 &              4.00  \\
    HD 95735 &   11:03:20.1 &   +35:58:11.5 &   7.52 &  3.34 &    $3464 \pm 15^{13}$ &      IGRINS &      2015-01-23 &              2.00  \\
    BS 5019 &   13:18:24.3 &   -18:18:40.3 &   4.74 &  2.96 &     $5558 \pm 19^{2}$ &      IGRINS &      2015-01-06 &              6.00  \\
  HD 119850 &   13:45:43.7 &   +14:53:29.4 &   8.50 &  4.41 &    $3618 \pm 31^{13}$ &      IGRINS &      2015-01-27 &              2.00  \\
  HD 122120 &   13:59:19.4 &   +22:52:11.1 &   9.04 &  6.24 &     $4580 \pm 7^{11}$ &      IGRINS &      2015-01-27 &              6.00  \\
  HD 122652 &   14:02:31.6 &   +31:39:39.0 &   7.15 &  5.88 &     $6093 \pm 44^{5}$ &      IGRINS &      2015-01-27 &              4.00  \\
     GJ 570A &   14:57:28.0 &   -21:24:55.7 &   5.72 &  3.10 &    $4507 \pm 58^{13}$ &      IGRINS &      2014-05-27 &              1.33  \\
     GJ 576 &   15:04:53.5 &   +05:38:17.1 &   9.81 &  6.47 &   $4450 \pm 100^{15}$ &      IGRINS &      2015-01-27 &              6.00  \\
     GJ 758 &   19:23:34.0 &   +33:13:19.0 &   6.36 &  4.49 &     $5453 \pm 44^{5}$ &      IGRINS &      2014-10-10 &              3.00  \\
    GJ 820 A &   21:06:53.9 &   +38:44:57.9 &   5.21 &  2.68 &    $4361 \pm 17^{13}$ &      IGRINS &      2014-12-05 &              0.67  \\
    GJ 820 B &   21:06:55.2 &   +38:44:31.4 &   6.03 &  2.32 &    $3932 \pm 25^{13}$ &      IGRINS &      2014-12-05 &              0.67  \\
  HD 220339 &   23:23:04.8 &   -10:45:51.2 &   7.80 &  5.59 &     $5029 \pm 52^{2}$ &      IGRINS &      2014-12-07 &             13.00  \\
 HIP 117473 &   23:49:12.5 &   +02:24:04.4 &   8.99 &  5.04 &     $3646 \pm 60^{7}$ &      IGRINS &      2014-11-24 &              4.00  \\
    HD 4614 &   00:49:06.2 &   +57:48:54.6 &   3.44 &  1.99 &      $5973 \pm 8^{3}$ &        TS23 &      1998-07-16 &              2.50  \\
    HD 10700 &   01:44:04.0 &   -15:56:14.9 &   3.50 &  1.68 &     $5290 \pm 39^{3}$ &        TS23 &      1998-07-16 &              3.00  \\
      GJ 74 &   01:46:38.7 &   +12:24:42.3 &   8.89 &  6.32 &    $4638 \pm 72^{12}$ &        TS23 &      2008-04-12 &             20.00  \\
    HD 22049 &   03:32:55.8 &   -09:27:29.7 &   3.73 &  1.67 &    $5077 \pm 35^{13}$ &        TS23 &      2000-09-22 &              1.67  \\
    HR 1287 &   04:10:49.8 &   +26:28:51.4 &   5.40 &  4.48 &     $6912 \pm 80^{8}$ &        TS23 &      2008-03-30 &              5.00  \\
    HD 30652 &   04:49:50.4 &   +06:57:40.5 &   3.19 &  2.05 &     $6414 \pm 19^{3}$ &        TS23 &      1998-11-03 &              1.00  \\
   HD 40590 &   05:59:51.5 &   +00:03:21.4 &   8.07 &  6.91 &     $6528 \pm 75^{8}$ &        TS23 &      2004-02-03 &             21.67  \\
    HR 3538 &   08:54:17.9 &   -05:26:04.0 &   6.00 &  4.46 &     $5788 \pm 23^{2}$ &        TS23 &      2000-01-15 &             15.00  \\
     GJ 380 &   10:11:22.1 &   +49:27:15.2 &   6.61 &  3.26 &    $4085 \pm 14^{13}$ &        TS23 &      2012-10-02 &             13.33  \\
     GJ 411 &   11:03:20.1 &   +35:58:11.5 &   7.52 &  3.34 &    $3464 \pm 15^{13}$ &        TS23 &      2008-03-27 &             10.00  \\
     61 Vir &   13:18:24.3 &   -18:18:40.3 &   4.74 &  2.96 &     $5558 \pm 19^{2}$ &        TS23 &      2000-01-12 &             12.00  \\
     70 Vir &   13:28:25.8 &   +13:46:43.6 &   4.97 &  3.24 &     $5406 \pm 64^{3}$ &        TS23 &      1998-07-14 &              8.00  \\
    HD 142860 &   15:56:27.1 &   +15:39:41.8 &   3.84 &  2.62 &     $6222 \pm 13^{3}$ &        TS23 &      1998-07-14 &              2.50  \\
     GJ 699 &   17:57:48.4 &   +04:41:36.2 &   9.51 &  4.52 &    $3222 \pm 10^{13}$ &        TS23 &      2000-05-24 &             35.00  \\
   70 Oph A &   18:05:27.3 &   +02:29:59.3 &   4.20 &  1.79 &    $5407 \pm 52^{13}$ &        TS23 &      1998-07-14 &              3.00  \\
   16 Cyg A &   19:41:48.9 &   +50:31:30.2 &   5.95 &  4.43 &     $5750 \pm 57^{3}$ &        TS23 &      2005-10-12 &              6.67  \\
   16 Cyg B &   19:41:51.9 &   +50:31:03.0 &   6.20 &  4.65 &     $5678 \pm 66^{3}$ &        TS23 &      2002-09-21 &             13.33  \\
   61 Cyg B &   21:06:55.2 &   +38:44:31.4 &   6.03 &  2.32 &    $3932 \pm 25^{13}$ &        TS23 &      1998-07-14 &             10.00  \\
     GJ 864 &   22:36:09.6 &   -00:50:30.0 &   9.92 &  6.16 &     $3916 \pm 61^{7}$ &        TS23 &      2002-11-22 &             25.00  \\
  HD 216625 &   22:54:07.4 &   +19:53:31.3 &   7.02 &  5.73 &     $6212 \pm 44^{5}$ &        TS23 &      2001-07-25 &             20.00  \\

\enddata

\tablecomments{The temperatures come from the following sources, and are labeled as superscripts after the temperature. [1]:  \cite{Woolf2005}; [2]: \cite{Sousa2008}; [3]: \cite{Boyajian2013}; [4]: \cite{Alonso1996}; [5]: \cite{Valenti2005}; [6]: \cite{Neves2014}; [7]: \cite{Mann2015}; [8]: \cite{Casagrande2011}; [9]: \cite{Ramirez2005}; [10]: \cite{Casagrande2008}; [11]: \cite{Mishenina2012}; [12]: \cite{Casagrande2010}; [13]: \cite{Boyajian2012}; [14]: \cite{Pecaut2013}; [15]: \cite{Zboril1998}.
}

\end{deluxetable*}

\newpage
\clearpage
\LongTables
\begin{deluxetable*}{lcccccrccc}
\tabletypesize{\footnotesize}
\tablewidth{0pt}
\tablecaption{ Known Binary Stars \label{tab:known}}
\tablehead{
\colhead{} & \colhead{} & \colhead{} & \colhead{}& \colhead{} & \colhead{} & \colhead{}  & \colhead{} & \colhead{Exp. Time} \\
\colhead{Star} & \colhead{RA} & \colhead{DEC} & \colhead{SpT}& \colhead{V} & \colhead{K} & \colhead{Instrument}  & \colhead{Date} & \colhead{(min)}}

\startdata

   HIP 1366 &   00:17:5.50 &  +38:40:53.89 &         A2V & 4.62 &     4.42 &       TS23 &  2013-10-20 &   32.17 \\
   HIP 3300 &   00:42:3.90 &  +50:30:45.09 &         B2V & 4.80 &     5.08 &       TS23 &  2013-01-07 &   41.49 \\
  HIP 12719 &  02:43:27.11 &  +27:42:25.72 &         B3V & 4.64 &     4.97 &       TS23 &  2013-10-18 &   36.43 \\
  HIP 13165 &  02:49:17.56 &  +17:27:51.52 &         B6V & 5.31 &     5.41 &        HRS &  2013-08-14 &   13.75 \\
  HIP 15338 &  03:17:47.35 &  +44:01:30.08 &         B8V & 5.48 &     5.59 &        HRS &  2013-08-19 &   28.50 \\
  HIP 17563 &  03:45:40.44 &  +06:02:59.98 &         B3V & 5.33 &     5.59 &     CHIRON &  2013-09-03 &  126.93 \\
  HIP 22840 &  04:54:50.71 &   +00:28:1.81 &         B5V & 5.97 &     6.25 &       TS23 &  2014-01-21 &   96.21 \\
  HIP 22958 &  04:56:24.19 &  -05:10:16.87 &         B6V & 5.49 &     5.79 &     CHIRON &  2013-09-16 &  140.00 \\
  HIP 22958 &  04:56:24.19 &  -05:10:16.87 &         B6V & 5.49 &     5.79 &     CHIRON &  2014-10-13 &   32.50 \\
  HIP 24902 &  05:20:14.67 &  +41:05:10.35 &         A3V & 5.47 &     5.02 &     IGRINS &  2014-10-16 &    3.73 \\
  HIP 26063 &  05:33:31.45 &  -01:09:21.87 &         B1V & 5.38 &     5.86 &     CHIRON &  2013-10-17 &  132.88 \\
  HIP 26563 &  05:38:53.08 &  -07:12:46.18 &         A4V & 4.80 &     4.42 &       TS23 &  2014-01-20 &   69.49 \\
  HIP 28691 &  06:03:27.37 &  +19:41:26.02 &         B8V & 5.13 &     5.36 &       TS23 &  2013-01-06 &   66.53 \\
  HIP 33372 &  06:56:25.83 &  +09:57:23.67 &        B8Vn & 5.91 &     6.08 &       TS23 &  2014-01-21 &  110.74 \\
  HIP 33372 &  06:56:25.83 &  +09:57:23.67 &        B8Vn & 5.91 &     6.08 &     IGRINS &  2014-10-17 &    5.33 \\
  HIP 44127 &  08:59:12.45 &  +48:02:30.57 &  A7V & 3.10 &     2.66 &       TS23 &  2014-01-20 &   18.24 \\
  HIP 58590 &  12:00:52.39 &  +06:36:51.56 &         A5V & 4.66 &     4.25 &     CHIRON &  2013-02-15 &   41.07 \\
  HIP 65477 &  13:25:13.54 &  +54:59:16.65 &   A5V & 4.01 &  \nodata &       TS23 &  2014-01-12 &   37.12 \\
  HIP 76267 &  15:34:41.27 &  +26:42:52.89 &        A1IV & 2.21 &     2.21 &     CHIRON &  2013-03-29 &    4.20 \\
  HIP 77516 &  15:49:37.21 &  -03:25:48.74 &         A0V & 3.55 &     3.70 &     CHIRON &  2013-03-29 &   14.70 \\
  HIP 77858 &  15:53:53.92 &  -24:31:59.37 &         B5V & 5.38 &     5.36 &     CHIRON &  2014-03-17 &   76.07 \\
  HIP 79199 &  16:09:52.59 &  -33:32:44.90 &         B8V & 5.50 &     5.65 &     CHIRON &  2014-03-18 &   49.49 \\
  HIP 79404 &  16:12:18.20 &  -27:55:34.95 &         B2V & 4.57 &     4.98 &     CHIRON &  2013-05-03 &   37.80 \\
  HIP 79404 &  16:12:18.20 &  -27:55:34.95 &         B2V & 4.57 &     4.98 &     CHIRON &  2015-02-23 &   21.78 \\
  HIP 79404 &  16:12:18.20 &  -27:55:34.95 &         B2V & 4.57 &     4.98 &     CHIRON &  2015-03-09 &   18.92 \\
  HIP 81641 &  16:40:38.69 &  +04:13:11.23 &         A1V & 5.77 &     5.74 &        HRS &  2013-04-22 &   16.00 \\
  HIP 84606 &  17:17:40.25 &  +37:17:29.40 &         A2V & 4.62 &     4.44 &     IGRINS &  2014-10-15 &    7.47 \\
  HIP 85385 &  17:26:49.13 &  +20:04:51.52 &         B5V & 5.51 &     5.84 &     IGRINS &  2014-07-10 &    8.00 \\
  HIP 88290 &  18:01:45.20 &  +01:18:18.28 &        A2Vn & 4.44 &     4.23 &     CHIRON &  2014-08-04 &   39.32 \\
  HIP 91118 &  18:35:12.60 &  +18:12:12.28 &        A0Vn & 5.79 &     5.67 &     IGRINS &  2014-10-15 &    6.00 \\
  HIP 92027 &  18:45:28.36 &   +05:30:0.44 &     A1V & 5.83 &     5.66 &        HRS &  2013-04-23 &   18.00 \\
  HIP 92728 &  18:53:43.56 &  +36:58:18.19 &       B2.5V & 5.57 &     5.99 &        HRS &  2013-04-23 &   14.00 \\
  HIP 98055 &  19:55:37.79 &  +52:26:20.21 &        A4Vn & 4.92 &     4.49 &       TS23 &  2013-10-21 &   42.82 \\
 HIP 100221 &  20:19:36.72 &  +62:15:26.90 &         B9V & 5.71 &     5.71 &        HRS &  2013-08-19 &   43.70 \\
 HIP 106786 &  21:37:45.11 &  -07:51:15.13 &         A7V & 4.69 &     4.25 &     CHIRON &  2014-05-17 &   23.75 \\
 HIP 106786 &  21:37:45.11 &  -07:51:15.13 &         A7V & 4.69 &     4.25 &     IGRINS &  2014-10-15 &    3.73 \\
 HIP 106786 &  21:37:45.11 &  -07:51:15.13 &         A7V & 4.69 &     4.25 &       TS23 &  2014-11-01 &   19.99 \\
 HIP 113788 &  23:02:36.38 &  +42:45:28.06 &        A3Vn & 5.10 &     4.69 &       TS23 &  2013-10-21 &   47.51 \\
 HIP 116247 &  23:33:16.62 &  -20:54:52.22 &         A0V & 4.71 &     4.52 &     CHIRON &  2013-06-20 &   42.93 \\
 HIP 116611 &  23:37:56.80 &   +18:24:2.40 &        A1Vn & 5.48 &     5.42 &     IGRINS &  2014-10-16 &    4.20 \\
 
\enddata
\tablecomments{The spectral types are from the Simbad database \citep{Simbad}.}
\end{deluxetable*}

\newpage
\clearpage
\begin{deluxetable}{cccc}
\tabletypesize{\footnotesize}
\tablecolumns{5}
\tablewidth{0pt}
\tablecaption{ Literature Spectroscopic Data \label{tab:specdata}}

\tablehead{
                           & \colhead{K1} & \colhead{K2} & \colhead{Period} \\
 \colhead{Star} &  \colhead{($\rm km \cdot s^{-1}$)}  &  \colhead{($\rm km \cdot s^{-1}$)} & \colhead{(days)} 
}

\startdata
   HIP 3300$^{4}$ &    11.90 &  \nodata &  940.20 \\
  HIP 12719$^{4}$ &     8.80 &  \nodata &  490.00  \\
  HIP 13165$^{5}$ &    24.80 &  \nodata &    3.85  \\
  HIP 15338$^{6}$ &    20.00 &  \nodata &   36.50 \\
  HIP 17563$^{7}$ &    26.80 &  \nodata &    1.69  \\
  HIP 22840$^{7}$ &    24.50 &  \nodata &   24.10  \\
  HIP 26063$^{11}$ &    13.50 &  \nodata &  119.09  \\
  HIP 26563$^{12}$ &    28.60 &  \nodata &  445.74  \\
  HIP 28691$^{13}$ &    12.22 &  \nodata & 4741.10  \\
  HIP 44127$^{12}$ &     6.00 &  \nodata & 4028.00  \\
  HIP 58590$^{12}$ &    26.20 &  \nodata &  282.69  \\
  HIP 76267$^{14}$ &    35.40 &    99.00 &   17.36  \\
  HIP 77858$^{15}$ &    32.90 &  \nodata &    1.92  \\
  HIP 79404$^{15}$ &    31.50 &  \nodata &    5.78  \\
  HIP 85385$^{7}$ &    17.10 &  \nodata &    8.96 \\
  HIP 92728$^{16}$ &    39.70 &  \nodata &   88.35  \\
 HIP 100221$^{18}$ &    49.70 &  \nodata &    5.30  \\
 HIP 106786$^{12}$ &    11.30 &  \nodata & 8016.00  \\
 HIP 116611$^{3}$ &    25.19 &  \nodata &    0.50  \\

\enddata
\tablecomments{Known binary stars with spectroscopic orbit solutions. The orbital data is from the SB9 database \citep{Pourbaix2004}, and the original references are provided as superscripts after the star names: [1]: \cite{Hill1971}; [2]: \cite{Lloyd1981}; [3]: \cite{Rucinski2005}; [4]: \cite{Abt1978} [5]: \cite{Pourbaix2004}; [6]: \cite{Morrell1992}; [7]: \cite{Abt1990}; [8]: \cite{Fekel1982}; [9]: \cite{Lucy1971}; [10]: \cite{Pogo1928}; [11]: \cite{Duerbeck1975}; [12]: \cite{Abt1965}; [13]: \cite{Scarfe2000}; [14]: \cite{Tomkin1986}; [15]: \cite{Levato1987}; [16]: \cite{Richardson1957}; [17]: \cite{Leone1999}; [18]: \cite{Hube1973};  [19]: \cite{Pearce1936}
}
\end{deluxetable}

\begin{deluxetable}{cccc}
\tabletypesize{\footnotesize}
\tablecolumns{5}
\tablewidth{0pt}
\tablecaption{ Literature Imaging Data \label{tab:imagedata}}

\tablehead{
 & \colhead{Separation} &  & \colhead{Wavelength} \\
 \colhead{Star} & \colhead{($''$)} & \colhead{$\Delta m$} & \colhead{(nm)}
}

\startdata

   HIP 1366$^{1}$ &            0.06 &          \nodata &    $549$  \\
  HIP 22958$^{3}$ &            0.65 &  $4.15 \pm 0.14$ &   $511$  \\
  HIP 24902$^{3}$ &            0.38 &  $2.97 \pm 0.06$ &   $511$  \\
  HIP 33372$^{3}$ &            0.75 &  $3.27 \pm 0.04$ &   $511$  \\
  HIP 65477$^{4}$ &            1.11 &  $5.18 \pm 0.07$ &  $4770$  \\
  HIP 77516$^{5}$ &            0.20 &  $1.70 \pm 0.05$ &     $780$  \\
  HIP 79199$^{6}$ &            1.12 &  $4.62 \pm 0.12$ &  $1250$  \\
  HIP 81641$^{7}$ &            0.04 &  $1.90 \pm 0.00$ &    $551$  \\
  HIP 84606$^{3}$ &            0.84 &  $4.02 \pm 0.08$ &   $511$  \\
  HIP 88290$^{3}$ &            0.58 &  $2.95 \pm 0.04$ &   $511$  \\
  HIP 91118$^{8}$ &            0.16 &          \nodata &    $549$  \\
  HIP 92027$^{9}$ &            0.17 &  $1.53 \pm 0.00$ &    $550$  \\
  HIP 98055$^{10}$ &            0.10 &  $0.51 \pm 0.00$ &    $550$  \\
 HIP 113788$^{3}$ &            0.39 &  $2.17 \pm 0.02$ &   $511$  \\
 HIP 116247$^{11}$ &            0.84 &  $2.43 \pm 0.15$ &    $541$  \\
 HIP 116611$^{12}$ &            0.95 &  $5.93 \pm 0.09$ &   $2169$  \\

\enddata
\tablecomments{Known binary stars detected through either high-contrast imaging or interferometry. The imaging data comes from the Washington Double Star Catalog \citep{WDS}, and the most recent measurements are given as superscripts to the star name: [1]: \cite{McAlister1989}; [2]: \cite{Roberts2007}; [3]: \cite{Hipparchos}; [4]: \cite{Mamajek2010}; [5]: \cite{Drummond2014}; [6]: \cite{Shatsky2002}; [7]: \cite{Tokovinin2010}; [8]: \cite{McAlister1987}; [9]: \cite{Horch2010}; [10]: \cite{Horch2008}; [11]: \cite{Horch2001}; [12]: \cite{DeRosa2012}
}
\end{deluxetable}

\newpage
\clearpage
\begin{deluxetable}{cccccccc}
\tabletypesize{\footnotesize}
\tablewidth{0pt}
\tablecaption{ Companion data \label{tab:measured}}
\tablehead{
\colhead{} &  \multicolumn{3}{c}{Measured Values} &  \multicolumn{2}{c}{Expected Values} \\
\colhead{ Primary} & \colhead{$\rm T_{eff}$} & \colhead{[Fe/H]} & \colhead{vsini} & \colhead{ $\rm T_{eff}$} & \colhead{ $v\sin{i}$} \\
\colhead{Star} & \colhead{ (K)} & \colhead{(dex)} & \colhead{ (km s$^{-1}$)} & \colhead{(K)} & \colhead{ (km s$^{-1}$)}}

\startdata

  HIP 13165 & $5770 \pm 162$ & -0.5  &  5 & \nodata & \nodata \\
  HIP 22958 & $6070 \pm 112$ & -0.5 &  30 &     $6240^{+579}_{-409}$ & $11^{+16}_{-8}$ \\
  HIP 24902 & $5680 \pm 154$ & 0.0  &  30 &      $5950^{+34}_{-84}$ & $4^{+3}_{-3}$ \\
  HIP 33372 &               \nodata &     \nodata &    \nodata &     $6955^{+238}_{-550}$ & $12^{+16}_{-8}$ \\
  HIP 65477 &               \nodata &     \nodata &    \nodata &       $3861^{+34}_{-11}$ & $2^{+3}_{-1}$ \\
  HIP 76267 &  $5450 \pm 158$ &   -0.5 &   5 &     $5670^{+193}_{-281}$ & $3^{+4}_{-2}$ \\
  HIP 77516 &  $6820 \pm 141$ &    0.5 &   5 &     $6600^{+530}_{-167}$ & $12^{+17}_{-8}$ \\
  HIP 79199 &  $4620 \pm 158$ &   -0.5 &   5 &     $4800^{+484}_{-484}$ & $6^{+7}_{-4}$ \\
  HIP 79404 &  $4770 \pm 112$ &   -0.5 &  10 &              \nodata &     \nodata \\
  HIP 81641 &               \nodata &     \nodata &    \nodata &     $6475^{+126}_{-96}$ & $10^{+12}_{-7}$ \\
  HIP 84606 &  $5480 \pm 154$ &   0.0 &  10 &     $5450^{+71}_{-45}$ & $3^{+3}_{-2}$ \\
  HIP 88290 &               \nodata &     \nodata &    \nodata &     $5847^{+41}_{-24}$ & $4^{+4}_{-3}$ \\
  HIP 91118 &  $6490 \pm 154$ &   -0.5 &   10 &              \nodata &     \nodata \\
  HIP 92027 &               \nodata &     \nodata &    \nodata &     $6752^{+315}_{-96}$ & $12^{+16}_{-8}$ \\
  HIP 98055 &               \nodata &     \nodata &    \nodata &     $7366^{+391}_{-113}$ & $13^{+18}_{-9}$ \\
 HIP 113788 &               \nodata &     \nodata &    \nodata &      $6276^{+34}_{-92}$ & $7^{+6}_{-4}$ \\
 HIP 116247 &               \nodata &     \nodata &    \nodata &     $6351^{+482}_{-218}$ & $8^{+10}_{-6}$ \\
 HIP 116611 &               \nodata &     \nodata &    \nodata &     $3842^{+103}_{-50}$ & $2^{+4}_{-1}$ \\

\enddata

\end{deluxetable}

\newpage
\clearpage
\bibliography{references}

\end{document}